\documentclass[preprint,3p,authoryear,times]{elsarticle}

\usepackage{lineno,hyperref}
\modulolinenumbers[5]

\usepackage{amsmath,graphicx,float, url}
\usepackage{amssymb}
\usepackage{latexsym}
\usepackage{booktabs}
\usepackage{epsfig}
\usepackage{tablefootnote}
\usepackage{caption}
\usepackage{subcaption}
\usepackage{xcolor}

\newcommand{\fzero}    {f_{0}}

\usepackage{siunitx}
\usepackage{etoolbox}
\usepackage{dcolumn}
\newcolumntype{d}[1]{D{.}{.}{#1}}
\renewcommand{\bfseries}{\fontseries{b}\selectfont}
\newrobustcmd{\B}{\bfseries}

\journal{Computer Speech and Language}

\begin{document}
\begin{frontmatter}
\title{Analysis of DNN Speech Signal Enhancement for Robust Speaker Recognition}


\author{Ond\v{r}ej Novotn\'y\corref{cor1} }
\ead{inovoton@fit.vutbr.cz}
\author{Old\v{r}ich Plchot\corref{}}
\author{Ond\v{r}ej Glembek}
\author{Jan ``Honza" \v{C}ernock\'{y}}
\author{Luk\'a\v{s} Burget}

\address{Brno University of Technology, Speech@FIT and IT4I Center of Excellence, \\
Bo\v{z}et\v{e}chova 2, 612 66 Brno, Czech Republic}
 
\cortext[cor1]{Corresponding author}

\begin{abstract} 
In this work, we present an analysis of a DNN-based autoencoder for speech enhancement, dereverberation and denoising. The target application is a robust speaker verification (SV) system. We start our approach by carefully designing a data augmentation process to cover wide range of acoustic conditions and obtain rich training data for various components of our SV system. We augment several well-known databases used in SV with artificially noised and reverberated data and we use them to train a denoising autoencoder (mapping noisy and reverberated speech to its clean version) as well as an x-vector extractor which is currently considered as state-of-the-art in SV. Later, we use the autoencoder as a preprocessing step for text-independent SV system. We compare results achieved with autoencoder enhancement, multi-condition PLDA training and their simultaneous use. We present a detailed analysis with various conditions of NIST SRE 2010, 2016, PRISM and with re-transmitted data. We conclude that the proposed preprocessing can significantly improve both i-vector and x-vector baselines and that this technique can be used to build a robust SV system for various target domains.

\end{abstract}
\end{frontmatter}

\newcolumntype{C}{>{\centering\arraybackslash}X}%
\newcolumntype{R}{>{\raggedright\arraybackslash}X}%

\section{Introduction}
In recent years, there have been many attempts to take advantage of neural
networks (NNs) in speaker verification (SV). They  slowly found their way into the state-of-the-art systems that are based on modeling the fixed-length utterance representations, such as i-vectors~\citep{DehakN_TASLP:2010}, by Probabilistic Linear Discriminant Analysis (PLDA)~\citep{prince:iccv:2007}.

Most of the efforts to integrate the NNs into the SV pipeline involved replacing or improving one or more of the components of an i-vector + PLDA system (feature extraction, calculation of sufficient statistics, i-vector extraction or PLDA classifier) with a neural network.  On the front-end level, let us mention  for example using NN bottleneck features (BNF) instead of conventional Mel Frequency Cepstral Coefficients~\citep[MFCC,][]{lozano_odyssey_2016} or simply concatenating BNF and MFCCs~\citep{Matejka:ICASSP:2016} which greatly improves the performance and increases system robustness. Higher in the modeling pipeline, NN acoustic models can be used instead of Gaussian Mixture Models (GMM) for extraction of sufficient statistics~\citep{Lei_icassp_2014} or for either complementing PLDA~\citep{Novoselov_interspeech_2015,Bhattacharya_SLT16} or replacing it~\citep{Ghahabi_icassp_2014}.

These lines of work have logically resulted in attempts to train a larger DNN directly for the SV task, i.e., binary classification of two utterances as a \emph{target} or a \emph{non-target} trial~\citep{heighold_icassp_2016, zhang_slt_2016, sreEndEnd:Snyder2016, rohdin:icassp:2018}. Such architectures are known as \emph{end-to-end} systems and have been proven competitive for text-dependent tasks~\citep{heighold_icassp_2016, zhang_slt_2016} as well as text-independent tasks with short test utterances and an abundance of training data~\citep{sreEndEnd:Snyder2016}. In text-independent tasks with longer utterances and moderate amount of training data, the i-vector inspired end-to-end system~\citep{rohdin:icassp:2018} already outperforms generative baselines, but at the cost of high complexity in memory and computational costs during training.

While the fully end-to-end SV systems have been struggling with large requirements on the amount of training data (often not available to the researchers) and high computational costs, focus in SV has shifted back to generative modeling, but now with utterance representations obtained from a single NN. Such NN takes the frame level features of an utterance as an input and directly produces an utterance level representation, usually referred to as an \emph{embedding}~\citep{Variani_icassp_2014, heighold_icassp_2016, zhang_slt_2016, Bhattacharaya_interspeech_2017, xvec:Snyder2016}. The embedding is obtained by the means of a \emph{pooling mechanism} (for example taking the mean) over the frame-wise outputs of one or more layers in the~NN~\citep{Variani_icassp_2014}, or by the use of a recurrent NN~\citep{heighold_icassp_2016}. 
One effective approach is to train the NN for classifying a set of training speakers, i.e., using multiclass training~\citep{Variani_icassp_2014, Bhattacharaya_interspeech_2017,xvec:Snyder2016}. In order to do SV, the embeddings are extracted and used in a standard backend, e.g., PLDA. Such systems have recently been proven superior to i-vectors for both short and long utterance durations in text-independent SV~\citep{xvec:Snyder2016, xvec:Snyder2018}. 

Hand in hand with development of new modeling techniques that increase the performance of SV on particular benchmarks comes a requirement to continuously verify stability and improve robustness of the SV system under various scenarios and acoustic conditions.
One of the most important properties of a robust system is the ability to cope with the distortions caused by noise and reverberation and by the transmission channel itself. 
In SV, one way is to tackle this problem in the late modeling stage and use multi-condition training~\citep{david:icassp:vts,lei:multistyle} of PLDA, where we introduce noise and reverberation
variability into the within-class variability of speakers. This approach can be further combined with domain adaptation \citep{glembek:domainadaptation} which requires having certain amount of usually unsupervised target data. In the very last stage of the system, SV outputs can be adjusted per-trial basis via various kinds of adaptive score normalization \citep{Sturim:snorm,matejka:snorm,niko:norm}.

Another way to increase the robustness is to focus on the quality of the input acoustic signal and enhance it before it enters the SV system. Several techniques were introduced in the field of microphone arrays, such as active noise canceling, beamforming and filtering~\citep{kumatani:micarray:2012}.
For single microphone systems, front-ends utilize signal pre-processing methods, for example Wiener filtering, adaptive voice activity detection (VAD), gain control, etc.~\cite{ETSI:07}.
Various designs of robust features~\citep{plchot:rats13} can also be used in combination with normalization techniques such as cepstral mean and variance normalization or short-time gaussianization~\citep{Pelecanos01}.

At the same time when DNNs were finding their way into basic components of the SV systems, the interest in NN has also increased in the field of signal pre-processing and speech enhancement. An example of classical approach to remove a room impulse response is proposed in~\cite{Dufera2009}, where the filter is estimated by an NN. NNs have also been used for speech separation in~\cite{Yanhui2014}. NN-based autoencoder for speech enhancement was proposed in~\cite{Xu2014} with optimization in~\cite{Xu2014a} and finally, reverberant speech recognition with signal
enhancement by a deep autoencoder was tested in the Chime Challenge
and presented in~\cite{Mimura2014}.  

In this work, we focus on improving the robustness of SV via a DNN autoencoder as an audio pre-processing front-end. The autoencoder is trained to learn a mapping from noisy and reverberated speech to clean speech. The frame-by-frame aligned examples for DNN training are artificially created by adding noise and reverberation to the Fisher speech corpus.  
Resulting SV systems are tested both on real and simulated data. The real data cover both telephone conversations (NIST SRE2010 and SRE2016) and speech recorded over various microphones (NIST SRE2010, PRISM, Speakers In The Wild - SITW). Simulated data are created to produce challenging conditions by either adding the noise and reverberation into the clean microphone data or by re-transmission of the clean telephone and microphone data to obtain naturally reverberated data. 

After we explore the benefits of DNN-based audio pre-processing with standard generative SV systems based on i-vectors and PLDA, we attempt to improve an already better baseline system where DNN replaces the crucial i-vector extraction step. We use the architecture proposed by David Snyder~\cite{kaldi:xvec}, \cite{xvec:Snyder2016} which already presents the \emph{x-vector} (the embedding) as a robust feature for PLDA modeling, and provides state-of-the-art results across various acoustic conditions~\citep{xvec:Novotny2018}. 
We experiment with using the denoising autoencoder as a pre-processing step while training the x-vector extractor or just during the test stage. To further compare with the best i-vector system, we also experiment with using SBN features concatenated with MFCCs to train our x-vector extractor.


Finally, we offer experimental evidence and thorough analysis to demonstrate that the DNN-based signal enhancement increases the performance of the text-independent speaker verification system for both i-vector and x-vector based systems. We further combine the proposed method with multi-condition training that can significantly improve the SV performance and we show that we can profit from combination of both techniques.


\section{Speaker Recognition Systems (SRE)}

In this work we compare four systems, combining two feature extraction 
techniques---MFCC, and Stack Bottle-neck 
features (SBNs) concatenated with MFCCs---and two front-end modelling techniques---the i-vectors 
and the x-vectors, defined in~\cite{Matejka:Odyssey2014},~\cite{PLDA:kenny},~\cite{DehakN_TASLP:2010} and~\cite{xvec:Snyder2016}.  Please note, that each of the 
modeling techniques uses slightly different MFCC extraction.  See further description for details.

After feature extraction, voice activity detection (VAD) was performed by the 
BUT Czech phoneme recognizer, described in~\cite{matejka-odyssey-2006}, dropping all frames 
that are labeled as silence or noise.  The recognizer was trained on Czech 
CTS data, but we have added noise with varying SNR to 30\% of the database.  This VAD
was used both in the hyper-parameter training, as well as in the testing phase.

In all cases, speaker verification score was produced by comparing two i-vectors
(or x-vectors) corresponding to the segments in the verification trial by
Probabilistic Latent Discriminant Analysis~\citep[PLDA,][]{PLDA:kenny} for 
reference].


%
%


\subsection{MFCC i-vector system}
\label{lab:mfccsre}

In this system, we used cepstral features, extracted using a 25\,ms Hamming
window.  We used 24 Mel-filter banks and we limited the bandwidth to the
120--3800Hz range.  19~MFCCs together with
zero-\emph{th} coefficient were calculated every 10\,ms. This
20-dimensional feature vector was subjected to short time mean- and
variance-normalization using a 3\,s sliding window.  Delta and double delta
coefficients were then calculated using a five-frame window, resulting in a
60-dimensional feature vector. 

The acoustic modelling in this system is based on i-vectors.
To train the i-vector extractor, we use 2048-component diagonal-covariance 
Universal Background Model (GMM-UBM), and we set the dimensionality of i-vectors
to 600.  We then apply LDA to reduce the dimensionality to 200.  Such i-vectors
are then centered around a global mean followed by length 
normalization~\citep{DehakN_TASLP:2010,RomeroD_ICSLP:2011}.


\subsection{SBN-MFCC i-vector system}
\label{lab:bnmfccsre}



\begin{figure}[tb]
\centering\includegraphics[width=1.0\linewidth]{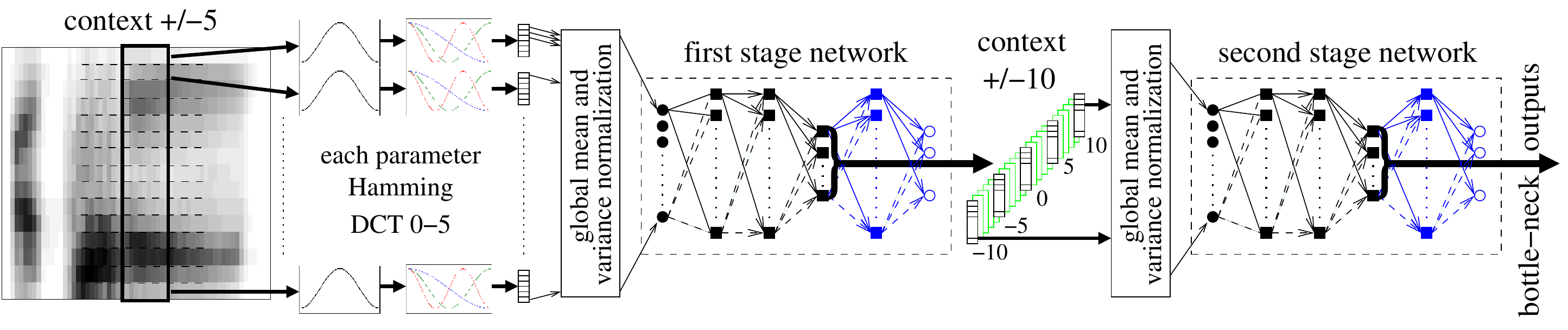}
\caption{Block diagram of Stacked Bottle-Neck (SBN) feature extraction. The blue parts of neural networks are used only during the training. The green frames in context gathering between the two stages are skipped. Only frames with shift -10, -5, 0, 5, 10 form the input to the second stage NN.}
\label{fig:BNfea}
\end{figure}

Bottleneck Neural-Network (BN-NN) refers to such a topology of a NN, where one 
of the hidden layers has significantly lower dimensionality than the surrounding
ones.
A bottleneck feature vector is generally understood as a by-product of forwarding
a primary input feature vector through the BN-NN and reading off the vector of
values at the bottleneck layer.
We have used a cascade of two such NNs for our experiments.
The output of the first network is
\emph{stacked} in time, defining context-dependent input features for the
second NN, hence the term Stacked Bottleneck features (Figure \ref{fig:BNfea}).

The NN input features are 24 log Mel-scale filter bank outputs
augmented with fundamental frequency features from 4 different $\fzero$
estimators (Kaldi,
Snack\footnote{http://kaldi.sourceforge.net, http://www.speech.kth.se/snack/}, and other two according to~\cite{Laskowski:LREC:2010} and~\cite{talkin:pitch:95}).
Together, we have 13 $\fzero$ related features, see~\cite{Karafiat:IS2014} for
details.
Conversation-side based mean subtraction is applied on the whole
feature vector, then 11 frames of log filter bank outputs and fundamental
frequency features are stacked. Hamming window and 
DCT projection (0$^{th}$ to 5$^{th}$ DCT base)  are applied on the time
trajectory of each parameter resulting in $(24+13)\times 6 = 222$
coefficients on the first stage NN input.

The configuration of the first NN is $222\times D_{H}\times D_{H}\times
D_{BN}\times D_{H}\times K$, where $K=9824$ is the number of target triphones. The
dimensionality of the bottleneck layer, $D_{BN}$ was set to 30.
The dimensionality of other hidden layers $D_{H}$ was set to 1500.
The bottleneck outputs from the first NN are sampled at times
$t{-}10$, $t{-}5$, $t$, $t{+}5$ and $t{+}10$, where $t$ is the index
of the current frame. The resulting 400-dimensional features are inputs to the
second stage NN with the same topology as the first stage. The network was trained on Fisher English corpus, and data were augmented with two noisy copies.

Finally, the 30-dimensional bottleneck outputs from the second NN (referred to as SBN) were concatenated with MFCC features (as used in the previous system) and used as an input to the conventional GMM-UBM i-vector system, with 2048 components in the UBM and 600-dimensional i-vectors.


\subsection{The x-vector systems}
\label{xvecdescription}

These SRE systems are based on a deep neural network (DNN) architecture for the extraction of
embeddings as described in~\cite{xvec:Snyder2016} and~\cite{xvec:Snyder2018}. 
Specifically, we use the original Kaldi recipe~\citep{kaldi:xvec} and 512 dimensional 
embeddings extracted from the first layer after the pooling layer (embedding-a, also 
referred to as the x-vector), which is consistent with~\cite{xvec:Snyder2018}.

Input features to the DNN were MFCCs, extracted using a 25\,ms Hamming window.  We used 23 Mel-filter banks and we limited the bandwidth to  20--3700\,Hz range.  23~MFCCs were calculated every 10\,ms.  This 20-dimensional feature vector was subjected to short time mean- and variance-normalization using a 3\,s sliding window.  Note the differences to the MFCC features for i-vector system described above (mainly the number of Mel-filter 
banks,  bandwidth, no delta/double delta coefficients).



The embedding DNN can be divided into three parts. The first part operates on the frame level and begins with 5 layers of time-delay architecture, described in~\cite{Peddinti:interspeech:2015}. The first four layers contain each 512 neurons, the last layer before statistics pooling has 1500 neurons.  The consequent pooling layer gathers mean and standard deviation statistics from all frame-level inputs. The single vector of concatenated means and standard deviations is  propagated through the rest of the network, where embeddings are extracted. This part consists of two hidden layers each with 512 neurons and the final output layer has a dimensionality corresponding to the number of speakers. The DNN uses Rectified Linear Units (ReLUs) as nonlinearities in hidden layers, soft-max in the output layer and is trained by optimizing multi-class cross entropy.

In addition, we also trained an x-vector extractor on MFCC features concatenated with SBN from Section~\ref{lab:bnmfccsre}.  Apart from changing the input features, we kept the architecture of the embedding DNN the same as for the MFCC system. 


\section{Signal Enhancement Autoencoder}
\label{ref:nnautoencoder}

\begin{figure}[tb]
    \begin{center}
      \scalebox{0.4}{\includegraphics{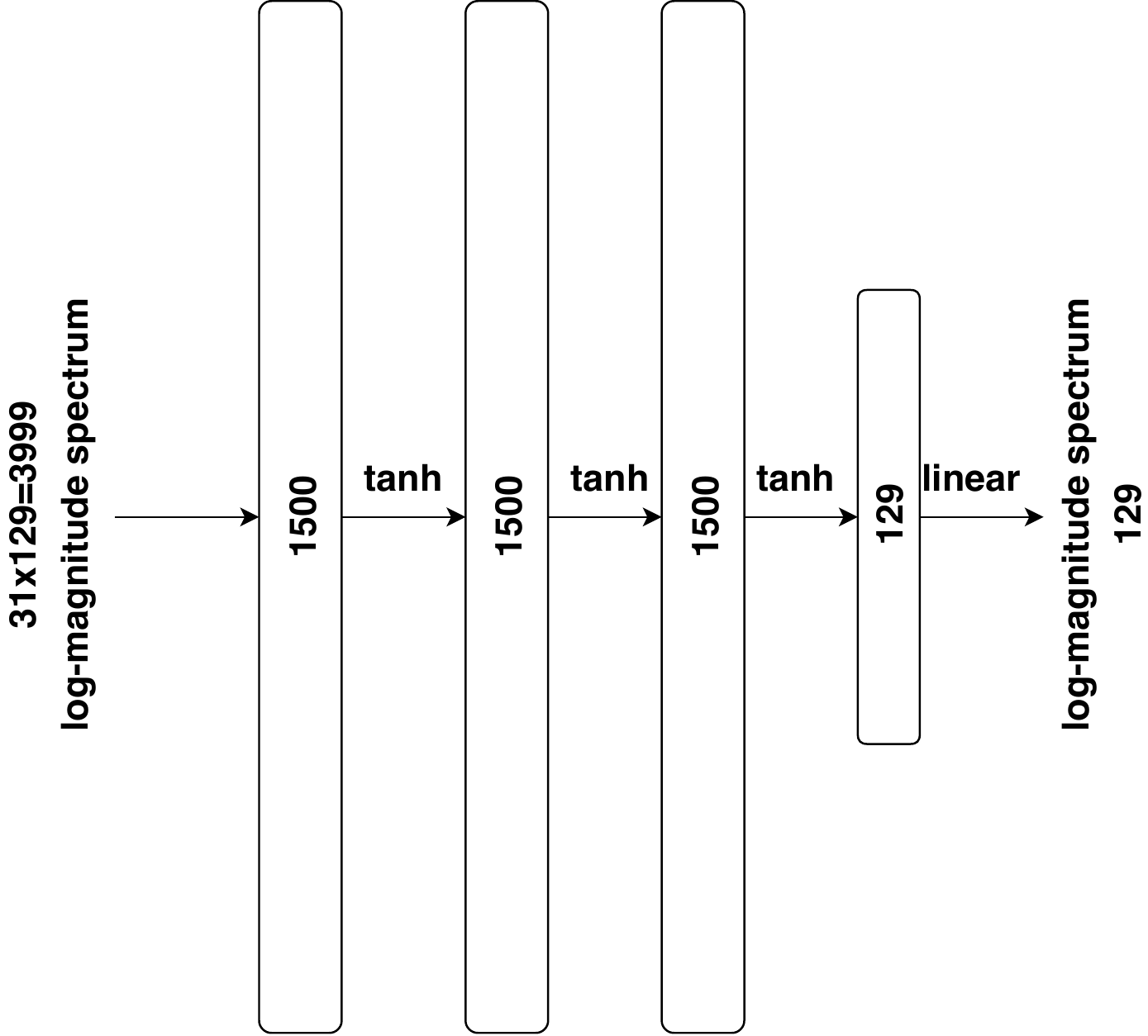}}
    \end{center}
    \vspace{-5mm}
    \caption{Topology of autoencoder:  three hidden layers each with 1500 neurons and hyperbolic tangent activation functions, output layer with 129 neurons and linear activation functions. The input of the network are 31 concatenated frames of the 129-dimensional log-magnitude spectrum.}
    \label{fig:autoencoder}
\end{figure}

For training the denoising autoencoder, we needed fairly large amount of clean speech from which we formed a parallel dataset of clean and augmented (noisy, reverberated or both) utterances. We chose Fisher English database Parts 1 and 2 as they span a large number of speakers (11971) and the audio is relatively clean and without reverberation. 
These databases combined contain over 20,000 telephone conversational sides or approximately 
1800 hours of audio.

Our autoencoder introduced in~\cite{plchot-enhancement-2016} and in~\cite{nn:Novotny2018} consists of three hidden layers with 1500 neurons in each layer.  The input of the autoencoder was a central frame of a log-magnitude spectrum with a context of +/- 15 frames (in total 3999-dimensional input). The output is a 129-dimensional enhanced central frame log-magnitude spectrum, see the topology in Figure~\ref{fig:autoencoder}.

It was necessary to perform feature normalization during the training and then repeat similar process during actual denoising. We used the mean and variance normalization with mean and variance estimated per input utterance. At the output layer, de-normalization with parameters estimated on a clean variant of the file was used during training while during denoising, the mean and variance were global and estimated on the cross-validation set. Using log on top of the magnitude spectrum decreases the dynamic range of the features and leads to a faster convergence.

As an objective function for training the autoencoder, we used the Mean Square Error (MSE) between the autoencoder outputs from training utterances and spectra of their clean variants. We were using both clean and augmented recordings during the training as we wanted the autoencoder to keep its robustness and produce good results also on relatively clean data.

\subsection{Adding noise}
We prepared a dataset of noises that consists of three different sources:
\begin{itemize}
\item 240 samples (4 minutes long) taken from the Freesound library\footnote{\url{http://www.freesound.org}} (real fan, HVAC, street, city, shop, crowd, library, office and workshop).
\item 5 samples (4 minutes long) of artificially generated noises: various spectral modifications of white noise + 50 and 100 Hz hum.
\item 18 samples (4 minutes long) of babbling noises by merging speech from 100 random speakers from Fisher database using speech activity detector. 
\end{itemize}

\noindent
Noises were divided into two disjoint groups for training (223 files) and development (40 files).

\subsection{Reverberation}
We prepared a set of with room impulse responses (RIRs) consisting of real room impulse responses from several databases: AIR\footnote{http://www.iks.rwth-aachen.de/en/research/tools-downloads/databases/aachen-impulse-response-database/}, C4DM\footnote{http://isophonics.net/content/room-impulse-response-data-set}~\citep{C4DM:WWW2}, MARDY\footnote{http://www.commsp.ee.ic.ac.uk/~sap/resources/mardy-multichannel-acoustic-reverberation-database-at-york-database/}, OPENAIR\footnote{http://www.openairlib.net/auralizationdb}, RVB 2014\footnote{http://reverb2014.dereverberation.com/index.html}, RWCP\footnote{http://www.openslr.org/13/} and RVB 2014\footnote{http://reverb2014.dereverberation.com/index.html}.
Together, they cover all types of rooms (small rooms, big rooms, lecture room, restrooms, halls, stairs etc.). All room models have more than one impulse response per room (different RIR was used for source of the signal and source of the noise to simulate their different locations). 
Rooms were split into two disjoint sets, with 396 rooms for training and 40 rooms for development.


\subsection{Composition of the training set}
\label{section:signal_corruption}
\begin{figure}[tb]
    \begin{center}
      \scalebox{0.6}{\includegraphics{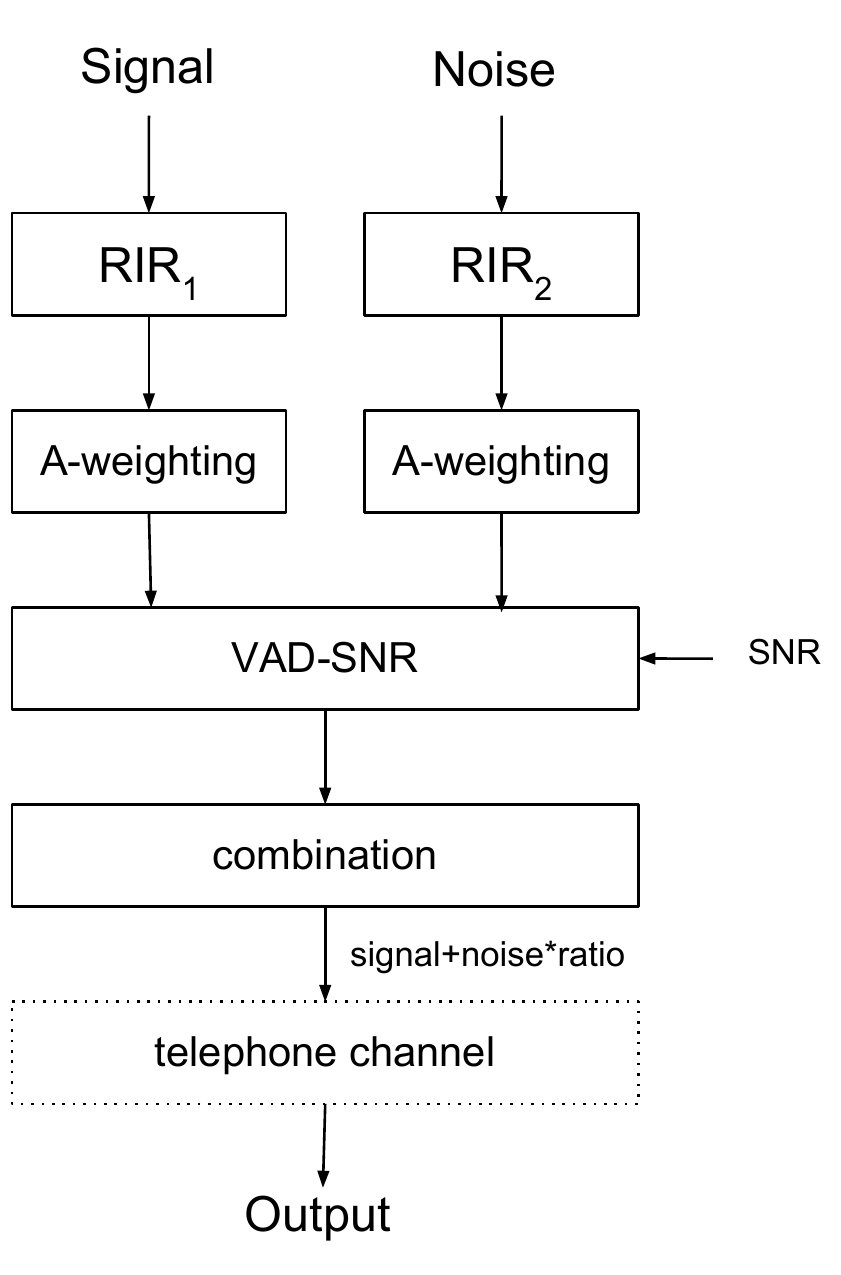}}
    \end{center}
    \vspace{-5mm}
    \caption{The process of data augmentation for autoencoder training, generating additional data for PLDA training, or system testing. The last step---filtering with the telephone channel---is used only when creating the denoising autoencoder training data.} \label{fig:noising_pipeline}
\end{figure}

To mix the reverberation, noise and signal at given SNR, we followed the  procedure showed in Figure~\ref{fig:noising_pipeline}.
The pipeline begins with two branches, when speech and noise are reverberated separately. Different RIRs from the same room are used for signal and noise, to simulate different positions of sources.

The next step is A-weighting, applied to simulate the perception of the human ear to added noise~\citep{aweighting}. With this filtering, the listener would be able to better perceive the SNR, because most of the noise energy is coming from frequencies, that the human ear is sensitive to. 

In the following step, we set a ratio of noise and signal energies to obtain the required SNR. Energies of the signal and noise are computed from frames given by original signal's voice activity detection (VAD). It means the computed SNR is really present in speech frames which are important for SV (frames without voice activity are removed during processing).

The useful signal and noise are then summed at desired SNR, and filtered with telephone channel \citep[see page 9 in][]{telfilter} to compensate for the fact that our noise samples are not coming from the telephone channel, while the original clean data (Fisher) are in fact telephone.
The final output is a reverberated and noisy signal with required SNR, which simulates a recording passing through the telephone channel (as was the original signal) in various acoustic environments. 
In case we want to add only noise or reverberation only, the appropriate part of the algorithm is used. 

\section{Experimental Setup}


\subsection{Training data}

To train the UBM and the i-vector extractor, we used the PRISM~\citep{ferrer:sre11} training dataset definition without added 
noise or reverberation. The PRISM set comprises 
Fisher 1 and 2, Switchboard phase 2 and 3 and Switchboard cellphone phases 
1 and 2, along with a set of Mixer speakers. This includes the 66 held out 
speakers from SRE10~\citep[see Section~III-B5 of][]{ferrer:sre11}, and 965, 980, 485 and 310 speakers 
from SRE08, SRE06, SRE05 and SRE04, respectively. A total of 13,916 speakers 
are available in Fisher data and 1,991 in Switchboard data.
Four variants of gender-independent PLDA were trained: the first variant was trained on 
the clean training data only, 
while the training sets for the other variants were augmented with artificially 
added mix of different noises and reverberated data (this portion was based on $30\%$ of the clean training data, i.e. approximately 24k segments).  


\subsection{Evaluation data}

We evaluated our systems on the \emph{female} portions of the following
NIST SRE 2010~\citep{NIST_SRE:WWW} and PRISM conditions:
\begin{itemize}
 \item {\it tel-tel}: SRE 2010 extended telephone condition 
involving normal vocal effort conversational telephone speech in enrollment and 
test (known as ``condition 5'').
  \item {\it int-int}: SRE 2010 extended interview condition involving 
interview speech from different microphones in enrollment and test (known as 
``condition 2'').
  \item {\it int-mic}: SRE 2010 extended interview-microphone condition 
involving interview enrollment speech and normal vocal effort conversational 
telephone test speech recorded over a room microphone channel (known as 
``condition 4'').
 \item {\it prism,noi}: Clean and artificially noised waveforms from both interview and telephone conversations recorded over lavalier microphones.
 Noise was added  at different SNR levels and recordings are tested against each other.
 \item {\it prism,rev}: Clean and artificially reverberated waveforms from both interview and telephone conversations recorded over lavalier microphones.
Reverberation was added with different RTs and recordings are tested against each other.
\item {\it prism,chn}: English telephone conversation with normal vocal effort 
recorded over different microphones from both SRE2008 and 2010 are tested against each other.
\end{itemize}

Additionally, we used the {\it Core-Core} condition from the SITW challenge -- {\it sitw-core-core}. The SITW~\citep[see][]{SITW_evaluation_plan} dataset is a large collection of real-world data exhibiting speech from individuals across a wide array of challenging acoustic and environmental conditions. These audio recordings do not contain any artificially added noise, reverberation or other artifacts. This database was collected from open-source media. The {\it sitw-core-core} condition comprises audio files each containing a continuous speech segment from a single speaker. Enrollment and test segments contain between 6-180 seconds of speech. We evaluated all trials (both genders). 

We also tested our systems on the NIST SRE 2016, described in~\cite{NIST:SRE2016}, but we split the trial set by language into Tagalog ({\it sre16-tgl-f}) and Cantonese ({\it sre16-yue-f}).  We use only female trials (both single- and multi-session). Concerning the experiments with SRE'16, it is important to note that we did not use the SRE'16 unlabeled development set in any way, and we did not perform any score normalization (such as adaptive s-norm).

The speaker verification performance is evaluated in terms of the equal error rate  (EER).


\subsection{NIST retransmitted set (BUT-RET)}

To evaluate the impact of room acoustics on the accuracy of speaker verification, a proper dataset of reverberant audio is needed.
An alternative that fills a qualitative gap between unsatisfying simulation~\citep[despite the improvement of realism reported in][]{data-sim} and costly and demanding real speaker recording, is retransmission. 
To our advantage, we can also use the fact that a {\em known} dataset can be retransmitted so that the performances are readily comparable with known benchmarks. Hence, this was the method to obtain a new dataset.

\begin{figure}[tb]
  \centering
  \includegraphics[width=0.5\linewidth]{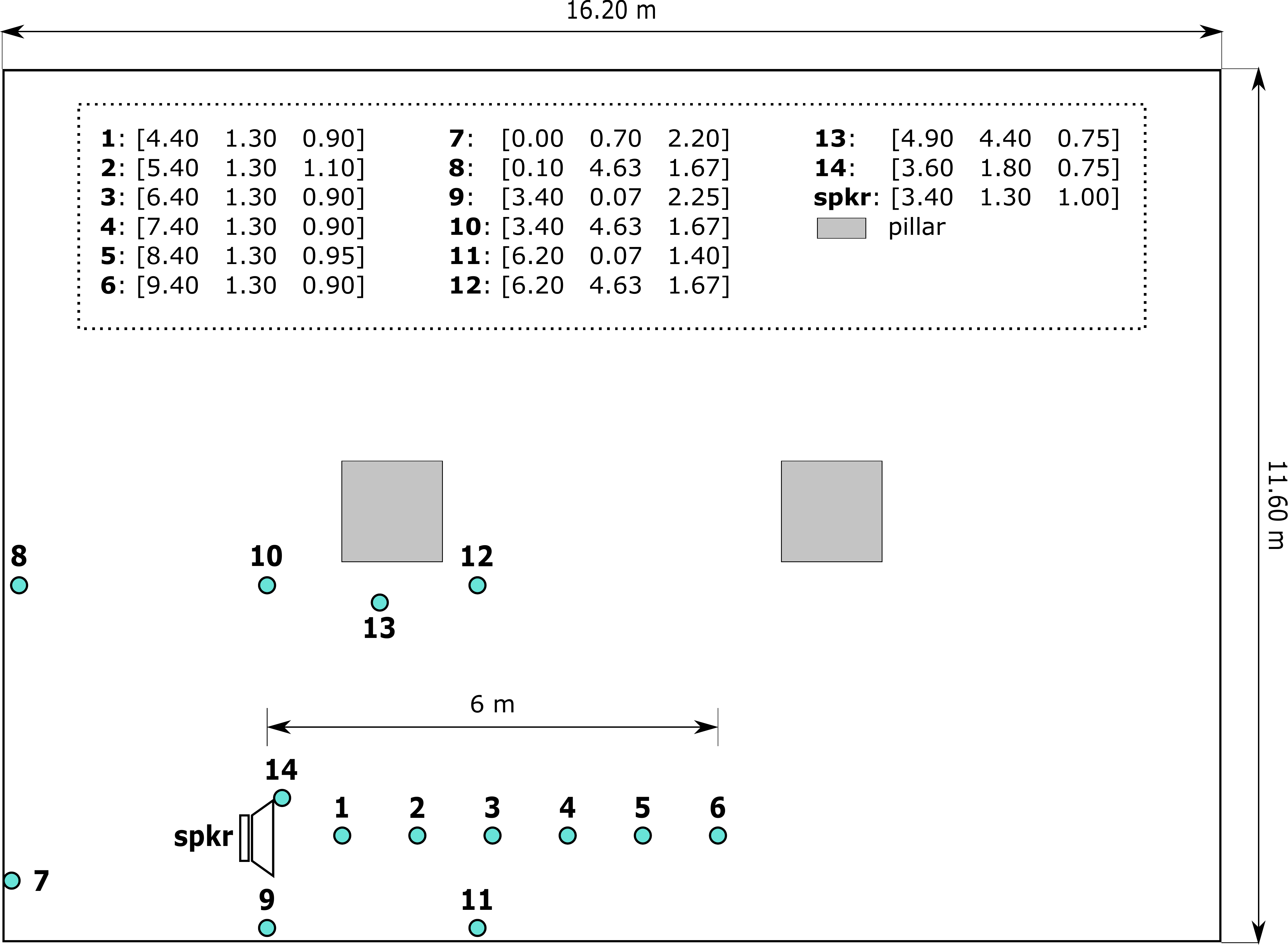}
  \caption{Floor plan of the room in which the retransmission took place. Coordinates are in meters and lower left corner is the origin.}
  \label{fig-floor-plan}
  \vspace{-5mm}
\end{figure}

The retransmission took place in a room with floor plan displayed in Figure~\ref{fig-floor-plan}. The configuration fits several purposes: the loudspeaker--microphone distance rises steadily for microphones 1\ldots{}6 to study deterioration as a function of distance, microphones 7\ldots{}12 form a large microphone array mainly focused to explore beamforming~\citep[beyond the scope of this paper but studied in][]{mosner:icassp2018}.

For this work, a subset of NIST SRE 2010  data 
was retransmitted. 
The dataset consists of 459 female recordings with nominal durations of three and eight minutes. The total number of female speakers is 150. The files were played in sequence and recorded simultaneously by a multi-channel acquisition card that ensured sample precision synchronization.


We denote the retransmitted data as condition \mbox{{\it BUT-RET-$*$}}, where \mbox{BUT-RET-{\it orig}}, represents original (not retransmitted) data and \mbox{BUT-RET-{\it merge}}, which is created by pooling scores from all fourteen microphones.


\subsection{PLDA augmentation sets}
\label{pldavariants}

For augmenting the PLDA training set, we created new artificially corrupted training sets from the PRISM training data. We used noises and RIRs described in Section~\ref{ref:nnautoencoder}. To mix the reverberation, noise and signal at given SNR, we followed the  procedure outlined in Figure~\ref{fig:noising_pipeline}, but omitting the last step of applying the telephone channel.
We trained the four following PLDAs (with abbreviations used further in the text):
\begin{itemize}
    \item \textbf{Clean}: PLDA was trained on original PRISM  data, without additive augmentation. 
    \item \textbf{N}:  PLDA was trained on i) original PRISM data, and ii) portion (24k segments) of the original training data corrupted by noise.
    \item \textbf{RR}:  PLDA was trained on i) original PRISM  data, and ii) portion of the original training data corrupted by reverberation using real room impulse responses.
    \item \textbf{RR+N}: PLDA was trained on i) original PRISM  data, ii) noisy augmented data, and iii) reverberated data as described above.
\end{itemize}
Note that the sizes of all 3 augmentation sets are the same.


\subsection{Augmentation sets for the embedding system}
\label{lab:nnaug}

When defining the data set for training the embedding system, we were trying to stay close to the recipe introduced by~\cite{kaldi:xvec}, but we introduced modifications to the training data that allowed us to test on larger set of benchmarks (PRISM, NIST SRE 2010).
Every speaker must have at least 6 utterances after augmentation (unlike 8 in the original recipe) and every training sample must be at least 500 frames long.  As consequence of these constraints and given the augmentation process described below, we ended up with 11383 training speakers. 

In the original Kaldi recipe, the training data were augmented with reverberation, noise, music, and babble noise and combined with the original clean data. The package of all noises and room impulse responses can be downloaded from OpenSLR\footnote{\url{http://www.openslr.org/resources/28/rirs_noises.zip}} 
\citep{RIR:Ko2017}, and includes MUSAN noise corpus (843 noises). 

For data {\em augmentation with reverberation}, the total amount of RIRs is divided into two equally distributed lists for medium and small rooms.  

For {\em augmentation with noise}, we created three replicas of the original data. The first replica was modified by adding MUSAN noises at SNR levels in the range of 0--15\,dB. In this case, the noise was added as a foreground noise (that means several non-overlapping noises can be added to the input audio).  The second replica was mixed with music at SNRs ranging from 5 to 15\,dB as background noise (one noise per audio with the given SNR). The last noisy replica of training data was created by mixing in babble noise. SNR levels were at 13--20\,dB and we used 3--7 noises per audio. The augmented data were pooled and a random subset of 200k audios was selected and combined with clean data. The process of data augmentation is also described in~\cite{xvec:Snyder2018}.

\begin{figure}[tb]
  \centering
  \includegraphics[width=0.55\linewidth]{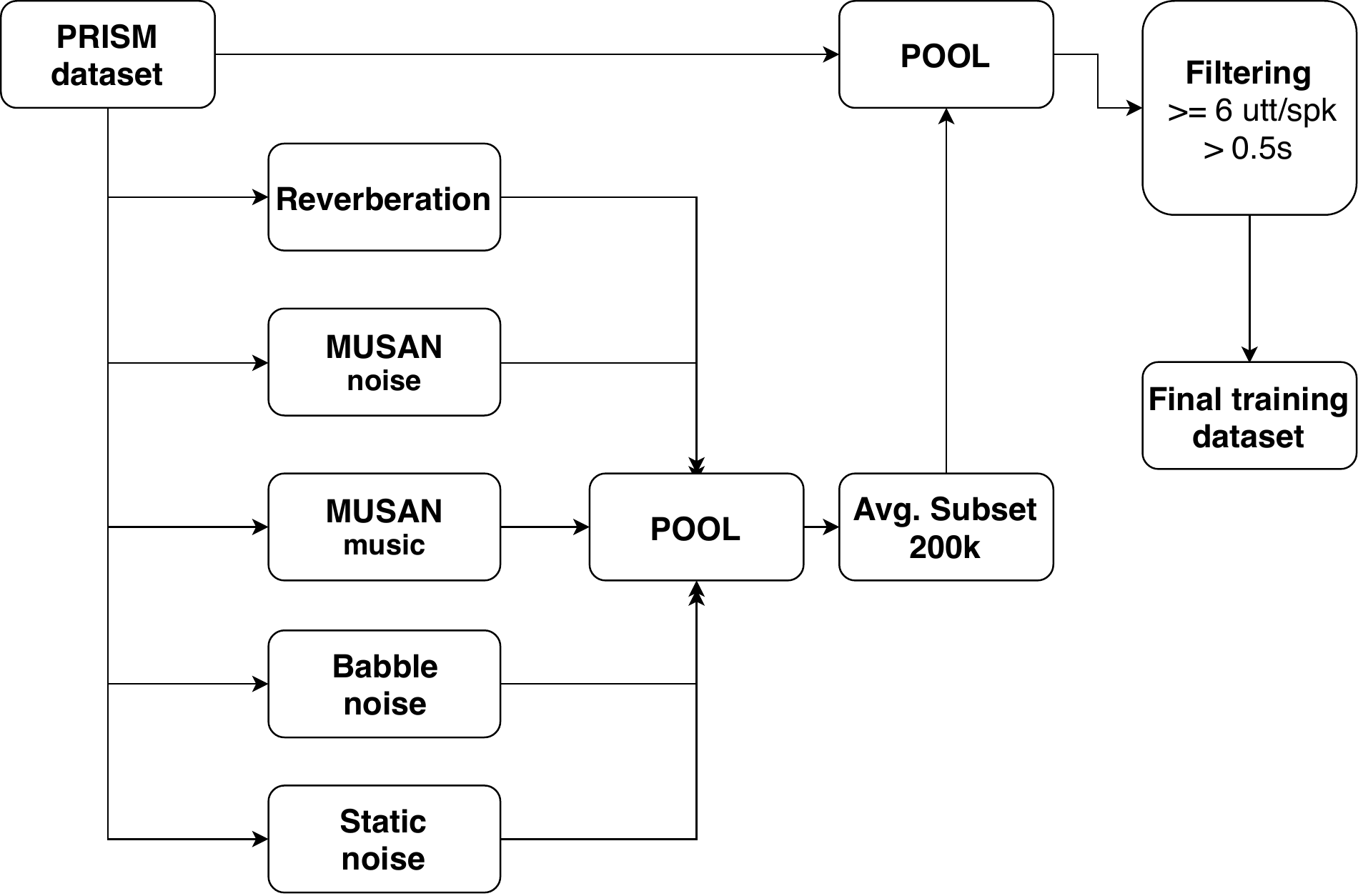}
  \caption{Data-flow diagram describing the preparation of the x-vector extractor training dataset.}
  \label{fig-kaldi}
\end{figure}

Apart from the original recipe, as described in the previous paragraph, we also added our own processing:  real room impulse responses and stationary noises described in Section~\ref{ref:nnautoencoder}.  
The original RIR list was extended by our list of real RIRs and we kept one reverberated replica. Our stationary noises were used to create another replica of data with SNR levels in range 0--20\,dB. We combined all replicas and selected a subset of 200k files. As a result, after performing all augmentations, we obtain 5 replicas for each original utterance. The whole process of creating the x-vector extractor training set is depicted in Figure~\ref{fig-kaldi}.

\section{Experiments and Discussion}
\label{exp}
We provide a set of results, where we study the influence of DNN autoencoder signal enhancement on a variety of systems.  Our autoencoder approach is also compared to the multi-condition training of PLDA, which can also improve the performance of the system in corrupted acoustic environment. At the end, we combine the autoencoder with the multi-condition training, and we find a better performing combination.

We trained autoencoders for signal enhancement simultaneously for denoising and dereverberation, which provides better robustness towards an unknown form of signal corruption, compared to autoencoder trained on noise or reverberation only \citep[as studied in][]{nn:Novotny2018}.

We also created different multi-condition training sets for PLDA (described in Section~\ref{pldavariants}),  similarly as for the autoencoder training (see Section~\ref{ref:nnautoencoder}). We used exactly the same noises and reverberation for segment corruption as in the autoencoder training, allowing to compare the performance of systems using the autoencoder and systems based on multi-condition training.  

Our results are listed in Table~\ref{tab:results-mfcc-ivec-compact} for the i-vector-based systems, and in Table~\ref{tab:results-mfcc-xvec-compact}  for the x-vector based ones.
The results in each table are separated into four main blocks based on a combination of features and signal augmentation: i) system trained with MFCC without signal enhancement, ii) system trained with MFCC with signal enhancement, iii) system trained with SBN-MFCC without enhancement, iv) and system trained with SBN-MFCC and signal enhancement.
In each block, the first column corresponds to the system where PLDA was trained only on clean data. The next three columns represent results when using different multi-condition training: N,  RR or N$+$RR  (as described in Section~\ref{pldavariants}).

Finally, the rows of the table are also divided based on the type of the condition, into telephone channel, microphone and artificially created conditions. The last row denoted as \emph{avg} gives the average EER over all conditions and each value set in bold is the minimum EER in the particular condition. We did not use any type of adaptation or any other technique used for results improvement in conditions from SRE16 and others. 


\subsection{I-vector systems experiments}

\begin{table*}[tbh]
\centering
\caption{\label{tab:results-mfcc-ivec-compact} Results (EER $[\%]$) obtained in four scenarios. Each block corresponds to a \textbf{i-vector} system trained with either MFCC or SBN-MFCC features and with or without signal enhancement applied during i-vector extraction.
Blocks are divided into columns corresponding to systems trained in multi-condition fashion (with noised and reverberated data in PLDA). Each column corresponds to a different PLDA multi-condition training set: ``---'' - clean condition, N - noise, RR - real reverberation, RR$+$N - real reverberation + noise. The last row denoted as \emph{avg} gives the average EER over all conditions and each value set in bold is the minimum EER in the particular condition.}
\scalebox{0.65}{
\begin{tabular}{ @{} l  S[table-format=2.2] S[table-format=2.2] S[table-format=2.2] S[table-format=2.2] c S[table-format=2.2] S[table-format=2.2] S[table-format=2.2] S[table-format=2.2] c S[table-format=2.2] S[table-format=2.2] S[table-format=2.2] S[table-format=2.2] c S[table-format=2.2] S[table-format=2.2] S[table-format=2.2] S[table-format=2.2] @{}}
\toprule
\multicolumn{1}{c}{} & \multicolumn{4}{c}{MFCC ORIG} & & 
                       \multicolumn{4}{c}{MFCC DENOISED} & &
                       \multicolumn{4}{c}{SBN-MFCC ORIG} & &
                       \multicolumn{4}{c}{SBN-MFCC DENOISED}\\ 

\cmidrule(rl){2-5} \cmidrule(rl){7-10} \cmidrule(rl){12-15} \cmidrule(rl){17-20}

Condition & 
\multicolumn{1}{c}{---} & \multicolumn{1}{c}{N}  & \multicolumn{1}{c}{RR} & \multicolumn{1}{c}{RR+N} & & 
\multicolumn{1}{c}{---} & \multicolumn{1}{c}{N}  & \multicolumn{1}{c}{RR} & \multicolumn{1}{c}{RR+N} & & 
\multicolumn{1}{c}{---} & \multicolumn{1}{c}{N}  & \multicolumn{1}{c}{RR} & \multicolumn{1}{c}{RR+N} & &
\multicolumn{1}{c}{---} & \multicolumn{1}{c}{N}  & \multicolumn{1}{c}{RR} & \multicolumn{1}{c}{RR+N}  \\ 

%
%
%
\midrule
tel-tel & 
    1.99 & 2.39 & 1.99 & 2.74 & &
    2.06 & 2.48 & 2.01 & 2.09 & &
    0.94 & 1.04 & 0.93 & 0.93 & &
    0.96 & 0.97 & 0.94 & $\mathbf{0.91}$ \\
sre16-tgl-f & 
    21.85 & 21.37 & 21.84 & 21.88 & &
    23.38 & 22.96 & 23.25 & 23.14 & &
    21.88 & $\mathbf{21.24}$ & 21.82 & 21.93 & &
    22.62 & 21.93 & 22.60 & 22.70 \\
sre16-yue-f & 
    11.20 & $\mathbf{10.52}$ & 11.15 & 11.53 & &
    11.76 & 11.47 & 11.76 & 11.79 & &
    13.45 & 13.02 & 13.45 & 13.44 & &
    14.60 & 13.69 & 14.54 & 14.52 \\
\midrule

int-int & 
    4.57 & 4.70 & 4.49 & 4.55 & &
    4.34 & 4.59 & 4.21 & 4.00 & &
    3.88 & 4.07 & 3.77 & 3.73 & &
    3.44 & 3.69 & $\mathbf{3.40}$ & $\mathbf{3.40}$ \\
int-mic & 
    1.85 & 2.09 & 1.86 & 2.00 & &
    2.51 & 2.33 & 2.40 & 2.32 & &
    1.85 & $\mathbf{1.69}$ & 1.76 & 1.78 & &
    1.87 & 1.79 & 1.84 & 1.77 \\
prism,chn & 
    1.03 & 1.29 & 0.99 & 0.97 & &
    0.59 & 0.67 & 0.59 & 0.57 & &
    0.40 & 0.46 & 0.39 & $\mathbf{0.36}$ & &
    0.66 & 0.78 & 0.72 & 0.66 \\
sitw-core-core & 
    10.11 & 10.13 & 10.06 & 10.32 & &
     9.41 &  9.60 &  9.45 &  9.45 & &
     8.09 &  7.85 &  8.02 &  8.03 & &
     7.71 &  $\mathbf{7.59}$ &  7.70 &  7.70 \\
\midrule
prism,noi & 
    3.72 & 3.02 & 3.65 & 3.42 & &
    2.51 & 2.38 & 2.46 & 2.38 & &
    2.43 & 1.98 & 2.45 & 2.20 & &
    1.84 & $\mathbf{1.73}$ & 1.81 & 1.76 \\
prism,rev & 
    2.51 & 2.67 & 2.40 & 2.23 & &
    1.94 & 2.09 & 1.89 & 1.92 & &
    1.42 & 1.39 & 1.30 & 1.31 & &
    1.12 & 1.23 & $\mathbf{1.07}$ & 1.09 \\
\midrule
BUT-RET-orig & 
    2.29 & 2.56 & 2.30 & 2.33 & &
    2.19 & 2.48 & 2.20 & 2.19 & &
    1.45 & 1.58 & 1.47 & $\mathbf{1.43}$ & &
    1.82 & 1.78 & 1.81 & 1.80 \\
BUT-RET-merge & 
    14.43 & 14.33 & 13.79 & 11.22 & &
    11.73 & 11.51 & 10.83 & 10.88 & &
    15.27 & 15.00 & 15.10 & 13.32 & &
     9.97 & 10.72 &  $\mathbf{9.38}$ &  9.47 \\

\midrule
avg & 
     6.87 & 6.82  & 6.77  & 6.65  & & 
     6.58 & 6.60  & 6.46  & 6.43  & & 
     6.46 & 6.30  & 6.41  & 6.22  & &
     6.06 & 5.99  & $\mathbf{5.98}$  & $\mathbf{5.98}$ \\

\bottomrule
\vspace{2mm}
\end{tabular}}
\end{table*}

Let us begin with comparing systems with and without signal enhancement. In this case, we  focus on PLDA trained on clean data only. 
In the first case, the i-vector system was trained using the MFCC features. We see mixed results. In the first set of conditions representing a telephone channel, we see degradation. When we consider that this is a reasonably clean condition, this enhancement was expected not to be very effective. 

In the second block of results (interview speech), the situation is better, except 
{\it int-mic} condition.  We can notice an improvement in the system with signal enhancement.  An interesting result can be spotted in condition {\it prism,chn}, where, with signal enhancement, we obtain more than 40\,\% relative improvement.  

The next block of artificially corrupted condition from PRISM also reports improvements and the last set of results with our retransmitted data too, in addition we can see there is no degradation in original condition {\it BUT-RET-orig}.

Let us now focus on the i-vector system based on the SBN-MFCC features. In the 
past, the SBN-MFCC features provided good robustness against noisy conditions. 
We verify  this statement comparing columns {\it MFFC-ORIG} and 
{\it SBN-MFCC-ORIG}  in  Table~\ref{tab:results-mfcc-ivec-compact} (systems without signal enhancement). We see that except for the SRE 2016 and {\it BUT-RET-merge} conditions, the system trained with stacked bottle-neck features yields better performance compared to the original MFCC system. 
When comparing systems with and without signal enhancement, the situation is
similar to the MFCC case. We see degradation on the telephone channels and a portion of the interview speech conditions. 
We obtain 30\,\% relative improvement in {\it BUT-RET-merge} where the system without enhancement is even worse than the previous i-vector system.  This could indicate that the bottle-neck features provide better robustness to noise than to reverberation.

In Section~\ref{lab:nnaug}, we described the augmentation setup for the x-vector system in comparison to the i-vector extractor training setup. Our presented i-vector extractors were trained on the original clean data only.  
Our hypothesis is that generative i-vector extractor training does not benefit from data augmentation in the same form as x-vector can.  
The comparison of our MFCC i-vector extractor trained on the original clean 
data and augmented data (the type of augmentation is the same as described in Section~\ref{ref:nnautoencoder}) is shown in
Table~\ref{tab:results-mfcc-ivec-iX-aug}. We see some improvement in some conditions, but mostly degradation. 
The reason is that generative i-vector extraction training is unsupervised.  When we add augmented data to the training list, i-vector extraction is forced to reserve a portion of parameters for representation of variability of noise, reverberation and so it limits parameters for speaker variability. 
In the supervised discriminative x-vector approach, we are forcing the x-vector extractor to do the opposite.  The extractor is forced to distinguish the speakers, and data augmentation in the training can be beneficial.

\begin{table*}[tbh]
\centering
\caption{\label{tab:results-mfcc-ivec-iX-aug} Results (EER $[\%]$) of i-vector extractor trained on clean data ({\it iX ORIG}) compared to i-vector extractor trained on augmented data ({\it iX AUG}).
Blocks are divided into columns corresponding to systems trained in multi-condition fashion (with noised and reverberated data in PLDA). Each column corresponds to a different PLDA multi-condition training set: ``---'' - clean condition, N - noise, RR - real reverberation, RR$+$N - real reverberation + noise.}
\scalebox{0.65}{
\begin{tabular}{ @{} l  
  S[table-format=2.2] S[table-format=2.2] S[table-format=2.2] S[table-format=2.2] 
c S[table-format=2.2] S[table-format=2.2] S[table-format=2.2] S[table-format=2.2] 
@{}
}
\toprule
\multicolumn{1}{c}{} & \multicolumn{4}{c}{iX ORIG} & & 
                       \multicolumn{4}{c}{iX AUG} \\ 

\cmidrule(rl){2-5} \cmidrule(rl){7-10}

Condition & 
\multicolumn{1}{c}{---} & \multicolumn{1}{c}{N}  & \multicolumn{1}{c}{RR} & \multicolumn{1}{c}{RR+N} & & 
\multicolumn{1}{c}{---} & \multicolumn{1}{c}{N}  & \multicolumn{1}{c}{RR} & \multicolumn{1}{c}{RR+N}  \\ 
\midrule
tel-tel & 
    1.99 & 2.39 & 1.99 & 2.74 & &
    $\mathbf{1.98}$ & 2.44 & 1.96 & 2.86 \\
sre16-tgl-f & 
    21.85 & $\mathbf{21.37}$ & 21.84 & 21.88 & &
    22.33 & 21.95 & 22.06 & 22.62 \\
sre16-yue-f & 
    11.20 & $\mathbf{10.52}$ & 11.15 & 11.53 & &
    11.32 & 10.59 & 11.26 & 11.20 \\
\midrule

int-int & 
    4.57 & 4.70 & 4.49 & 4.55 & &
    4.52 & 4.88 & $\mathbf{4.44}$ & 4.71 \\
int-mic & 
    $\mathbf{1.85}$ & 2.09 & 1.86 & 2.00 & &
    2.11 & 2.17 &  2.04 &  2.02 \\
prism,chn & 
    1.03 & 1.29 & 0.99 & 0.97 & &
    $\mathbf{0.92}$ & 1.20 &  0.95 &  1.04 \\
sitw-core-core & 
    10.11 & 10.13 & $\mathbf{10.06}$ & 10.32 & &
    10.28 & 10.38 & 10.17 &  10.34 \\
\midrule
prism,noi & 
    3.72 & $\mathbf{3.02}$ & 3.65 & 3.42 & &
    3.79 & 3.03 &  3.73 & 3.26 \\
prism,rev & 
    2.51 & 2.67 & 2.40 & 2.23 & &
    2.74 & 2.80 & 2.55 & $\mathbf{2.22}$ \\
\midrule
BUT-RET-orig & 
    $\mathbf{2.29}$ & 2.56 & 2.30 & 2.33 & &
    2.56 & 2.68 & 2.47 &  2.64 \\
BUT-RET-merge & 
    14.43 & 14.33 & 13.79 & 11.22 & &
    11.16 & 11.08 & 10.80 & $\mathbf{9.06}$  \\
\bottomrule
\vspace{2mm}
\end{tabular}}
\end{table*}

\subsection{X-vector systems experiments}

\begin{table*}[!tbh]
\centering
\sisetup{detect-weight, mode=text, table-format = 2.2}
\caption{\label{tab:results-mfcc-xvec-compact} Results (EER $[\%]$) obtained in four scenarios. Each block corresponds to an \textbf{x-vector} system trained with different type of features with or without signal enhancement.
Blocks are divided into columns corresponding to systems trained in multi-condition fashion (with noised and reverberated data in PLDA). Each column corresponds to different PLDA multi-condition training set: ``---'' - clean condition, N - noise, RR - real reverberation, RR$+$N - real reverberation + noise. The last row denoted as \emph{avg} gives the average EER over all conditions and each value set in bold is the minimum EER in the particular condition.}
\scalebox{0.65}{
\begin{tabular}{ @{} l  S[table-format=2.2] S[table-format=2.2] S[table-format=2.2] S[table-format=2.2] c S[table-format=2.2] S[table-format=2.2] S[table-format=2.2] S[table-format=2.2] c S[table-format=2.2] S[table-format=2.2] S[table-format=2.2] S[table-format=2.2] c S[table-format=2.2] S[table-format=2.2] S[table-format=2.2] S[table-format=2.2] @{}}
\toprule
\multicolumn{1}{c}{} & \multicolumn{4}{c}{MFCC ORIG} & & 
                       \multicolumn{4}{c}{MFCC DENOISED} & &
                       \multicolumn{4}{c}{SBN-MFCC ORIG} & &
                       \multicolumn{4}{c}{SBN-MFCC DENOISED}\\ 

\cmidrule(rl){2-5} \cmidrule(rl){7-10} \cmidrule(rl){12-15} \cmidrule(rl){17-20}


Condition & 
\multicolumn{1}{c}{---} & \multicolumn{1}{c}{N}  & \multicolumn{1}{c}{RR} & \multicolumn{1}{c}{RR+N} & & 
\multicolumn{1}{c}{---} & \multicolumn{1}{c}{N}  & \multicolumn{1}{c}{RR} & \multicolumn{1}{c}{RR+N} & & 
\multicolumn{1}{c}{---} & \multicolumn{1}{c}{N}  & \multicolumn{1}{c}{RR} & \multicolumn{1}{c}{RR+N} & &
\multicolumn{1}{c}{---} & \multicolumn{1}{c}{N}  & \multicolumn{1}{c}{RR} & \multicolumn{1}{c}{RR+N}  \\ 
\midrule

tel-tel & 1.30 & 1.43 & 1.27 & 1.29 & & 
          1.21 & 1.44 & $\mathbf{1.18}$ & 1.20 & &
          1.30 & 1.49 & 1.29 & 1.27 & &
          1.35 & 1.45 & 1.30 & 1.30 \\
sre16-tgl-f & 
        22.73 & 22.52 & 22.87 & 22.56 & &
        21.52 & 21.41 & 21.29 & 21.31 & &
        22.33 & 21.21 & 22.15 & 22.33 & &
        21.17 & $\mathbf{20.74}$ & 20.88 & 20.95 \\
sre16-yue-f & 
        10.36 & 9.61 & 10.45 & 10.61 & &
         8.86 & $\mathbf{8.23}$ &  8.75 &  8.66 & &
         9.60 & 8.71 &  9.56 &  9.88 & &
         8.89 & 8.38 &  8.67 &  8.64 \\

\midrule
int-int & 
    3.36 & 3.72 & 3.29 & 3.22 & &
    2.92 & 3.34 & 2.90 & $\mathbf{2.86}$ & &
    3.24 & 3.66 & 3.16 & 3.20 & &
    3.16 & 3.42 & 3.08 & 2.97 \\
int-mic & 
    1.33 & 1.43 & 1.3 & 1.22 & &
    1.47 & 1.37 & 1.41 & 1.37 & &
    1.07 & 1.17 & 1.04 & $\mathbf{1.03}$ & &
    1.37 & 1.39 & 1.29 & 1.27 \\
prism,chn & 
    0.62 & 0.81 & 0.61 & 0.61 & &
    0.37 & $\mathbf{0.27}$ & 0.36 & 0.41 & &
    0.69 & 0.67 & 0.67 & 0.59 & &
    0.36 & 0.41 & 0.36 & 0.36 \\
sitw-core-core & 
    7.87 & 7.30 & 7.72 & 7.41 & &
    6.81 & 6.54 & 6.73 & 6.70 & &
    7.57 & 7.42 & 7.57 & 7.38 & &
    6.81 & $\mathbf{6.40}$ & 6.71 & 6.66 \\
\midrule
prism,noi &
    2.76 & 1.90 & 2.63 & 2.11 & &
    1.84 & $\mathbf{1.50}$ & 1.80 & 1.73 & &
    2.72 & 1.97 & 2.63 & 2.22 & &
    1.84 & 1.57 & 1.81 & 1.68 \\
prism,rev & 
    2.08 & 2.02 & 1.79 & 1.60 & &
    1.16 & 1.13 & $\mathbf{1.10}$ & 1.12 & &
    1.98 & 2.06 & 1.71 & 1.59 & &
    1.24 & 1.27 & 1.15 & 1.15 \\
\midrule
BUT-RET-orig & 
    1.73 & 1.73 & 1.69 & 1.63 & &
    1.81 & 1.82 & 1.72 & 1.74 & &
    1.50 & 1.63 & 1.46 & {\textbf{ 1.43}} & &
    1.77 & 1.86 & 1.74 & 1.75 \\
BUT-RET-merge & 
    15.48 & 13.94 & 13.96 & 13.12 & &
    11.83 & 12.81 & $\mathbf{10.07}$ & 10.46 & &
    17.20 & 14.09 & 15.90 & 13.74 & &
    13.26 & 12.70 & 11.03 & 10.12 \\
\midrule
avg & 
     6.33 & 6.04 & 6.14 & 5.94 & &
     5.44 & 5.44 & 5.21 & 5.23 & &
     6.29 & 5.83 & 6.10 & 5.88 & &
     5.57 & 5.42 & 5.27 & {\textbf{ 5.17}} \\
     
\bottomrule
\vspace{2mm}
\end{tabular}}
\end{table*}

\begin{table*}[tbh]
\centering
\caption{\label{tab:results-xvec-ext-denoised} Results (EER $[\%]$) of SV system with x-vector extractor trained on clean data and with signal enhancement used only for x-vector extraction. Blocks are divided into columns corresponding to systems trained in multi-condition fashion (with noised and reverberated data in PLDA). Each column corresponds to a different PLDA multi-condition training set: ``---'' - clean condition, N - noise, RR - real reverberation, RR$+$N - real reverberation + noise.}
\scalebox{0.65}{
\begin{tabular}{ @{} l  
  S[table-format=2.2] S[table-format=2.2] S[table-format=2.2] S[table-format=2.2] 
c S[table-format=2.2] S[table-format=2.2] S[table-format=2.2] S[table-format=2.2] 
@{}
}
\toprule
\multicolumn{1}{c}{} & \multicolumn{4}{c}{MFCC} & & 
                       \multicolumn{4}{c}{SBN-MFCC} \\ 

\cmidrule(rl){2-5} \cmidrule(rl){7-10}

Condition & 
\multicolumn{1}{c}{---} & \multicolumn{1}{c}{N}  & \multicolumn{1}{c}{RR} & \multicolumn{1}{c}{RR+N} & & 
\multicolumn{1}{c}{---} & \multicolumn{1}{c}{N}  & \multicolumn{1}{c}{RR} & \multicolumn{1}{c}{RR+N}  \\ 
\midrule
tel-tel & 
    1.38 & 1.51 & 1.34 & 1.39 & &
    1.27 & 1.40 & $\mathbf{1.21}$ & 1.25 \\
sre16-tgl-f & 
    21.12 & 21.48 & 21.08 & $\mathbf{20.94}$ & &
    21.73 & 21.46 & 21.5 & 21.63 \\
sre16-yue-f & 
    9.76 & $\mathbf{9.01}$ & 9.7 & 9.69 & &
    9.38 & 9.07 & 9.41 & 9.16 \\
\midrule

int-int & 
    3.15 & 3.32 & 3.12 & $\mathbf{2.99}$ & &
    3.19 & 3.40 & 3.14 & 3.05 \\
int-mic & 
    1.61 & 1.67 & 1.59 & 1.58 & &
    1.63 & 1.58 & 1.51 & $\mathbf{1.39}$ \\
prism,chn & 
    0.54 & 0.47 & 0.55 & 0.54 & &
    $\mathbf{0.40}$ & 0.41 & $\mathbf{0.40}$ & $\mathbf{0.40}$ \\
sitw-core-core & 
    7.22 & 6.76 & 7.17 & 6.84 & &
    6.96 & $\mathbf{6.52}$ & 6.97 & 6.76 \\
\midrule
prism,noi & 
    2.14 & $\mathbf{1.64}$ & 2.15 & 2.05 & &
    2.33 & 1.67 & 2.36 & 2.15 \\
prism,rev & 
    1.24 & 1.22 & 1.18 & $\mathbf{1.20}$ & &
    1.33 & 1.45 & 1.28 & 1.24 \\
\midrule
BUT-RET-orig & 
    $\mathbf{1.87}$ & 2.07 & 1.9 & 1.88 & &
    2.09 & 2.03 & 2.08 & 2.07 \\
BUT-RET-merge & 
    12.76 & 11.76  & $\mathbf{10.71}$ & 11.83 & &
    15.08 & 14.32 & 12.62 & 12.66  \\
\bottomrule
\vspace{2mm}
\end{tabular}}
\end{table*}

We evaluated our speech enhancement autoencoder also with the system based on x-vectors, which is currently considered as state-of-the-art.  In our experiments and system design, we have deviated from the original Kaldi recipe~\citep{xvec:Snyder2018}. For training the x-vector extractor, we extended the number of speakers and we also created more variants of augmented data. We extended the original data augmentation recipe by adding real room impulse responses and an additional set of stationary noises (the extension process is also described in~\cite{xvec:Novotny2018}, the x-vector network used here is labeled as Aug III. in the paper). In the PLDA backend training, we also added the augmented data for multi-condition training (see Section~\ref{pldavariants}).

Let us point out, that the denoising autoencoder was trained on a subset of augmented data for training the x-vector DNN. The set of noises and real room impulse responses are therefore the same as in our extended set for training the x-vector extractor (as described in Section~\ref{ref:nnautoencoder}) and there is no advantage in autoencoder possibly seeing additional augmentations. 
It is also useful to refer the interested reader to our analysis in~\cite{xvec:Novotny2018}, where we show the benefit of having such a large augmentation set for x-vector extractor training.  

Let us first compare the x-vector network trained with original MFCC and with SBN-MFCC features.
In systems based on i-vectors, bottle-neck features provided sometimes very significant improvement, but for x-vector-based systems, the gains are much lower or the performance stays the same or even degrades for condition \emph{BUR-RET-merge}. This degradation, however, completely disappears after using denoising in x-vector training and subsequently multi-condition training in PLDA. For the telephone data with low reverberation, we can observe either steady performance on \emph{tel-tel} or slightly better performance on more challenging and non-English data in SRE'16 conditions. This is in contrast with i-vectors, where we only see either steady performance on easy \emph{tel-tel} or degradation on more challenging SRE'16. In general, the positive effect of SBN-MFCC features on x-vector system is small, but more stable than in i-vector system.

When we focus on the effect of signal enhancement in the x-vector-based system, we see much higher improvement compared to i-vectors. There are still several cases where the enhancement causes mostly degradation (MFCC: {\it int-mic}, {\it BUT-RET-orig}; SBN-MFCC: {\it tel-tel}, {\it int-mic}, {\it BUT-RET-orig}---mostly clean conditions). Otherwise, the enhancement provides nice improvement across rest of the conditions and features used for system training. At this point, it is useful to point out that unlike with i-vectors, where denoising is applied only for i-vector extraction, we actually apply enhancement already on top of x-vector training data. The effect of applying enhancement only during x-vector extraction like with i-vectors can be seen in Table~\ref{tab:results-xvec-ext-denoised}. We can observe that also here, we gain some improvements, but they are generally smaller than with enhancement deployed already during x-vector training (which can be observed in Table~\ref{tab:results-mfcc-xvec-compact}).

X-vector systems generally provide greater robustness across different signal corruptions. It was natural for us to expect, that x-vector systems should not need signal enhancement, and that they would implicitly learn it themselves, especially in the first part of DNN described in Section~\ref{xvecdescription}.  
To our belief, a reason why enhancement helped in our case is that denoising is not the target task of the x-vector DNN.  Even though we 
did have multiple corrupted samples per speaker in the DNN training set, 
it may be possible that we simply didn't have enough.  And since the x-vector
training is generally known to be data-hungry, it is therefore likely that if we had more corrupted samples per speaker, it would be in the DNN's natural capabilities to learn the task of de-noising.  


Let us also point out that if a single type of noise (or channel in general) appears systematically with a concrete speaker, the noise becomes a part of the speaker identity and therefore the NN does not compensate for it. 


So far, we have compared results on systems, where PLDA was trained on clean data only and we study possible improvements of enhancement across several systems.  Multi-condition training of PLDA, where we add a portion of augmented data into PLDA training is another possible approach on how to improve system performance and its robustness. 

From the results, we can see that multi-condition training, can provide improvement across all condition and systems without signal enhancement. We can see that the ideal combination of the augmented data for multi-condition training of PLDA depends on a condition. In noisy condition ({\it prism,noi}), it is
more effective to use noise augmentation only.  For reverberated condition ({\it prism,rev}, {\it BUT-RET-merge}) we can see more benefits in using reverberated augmentation set compared to others. 

\subsection{Final remarks}

\begin{figure}[p!]
\centering
\begin{subfigure}{.48\textwidth}
  \centering
    \centerline{
    \scalebox{1.0}{\includegraphics[width=0.5\linewidth]{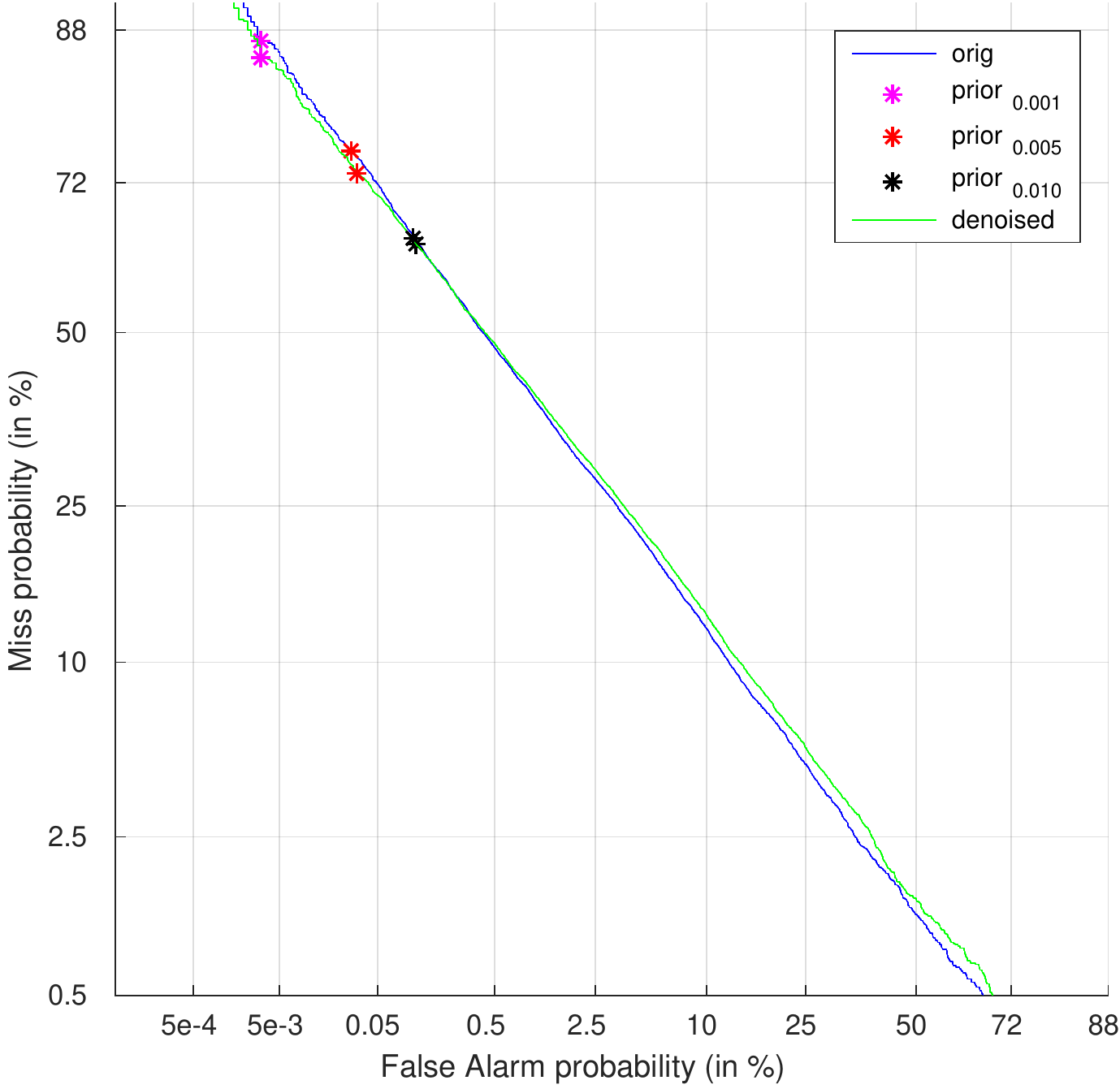}}
    \scalebox{1.0}{\includegraphics[width=0.5\linewidth]{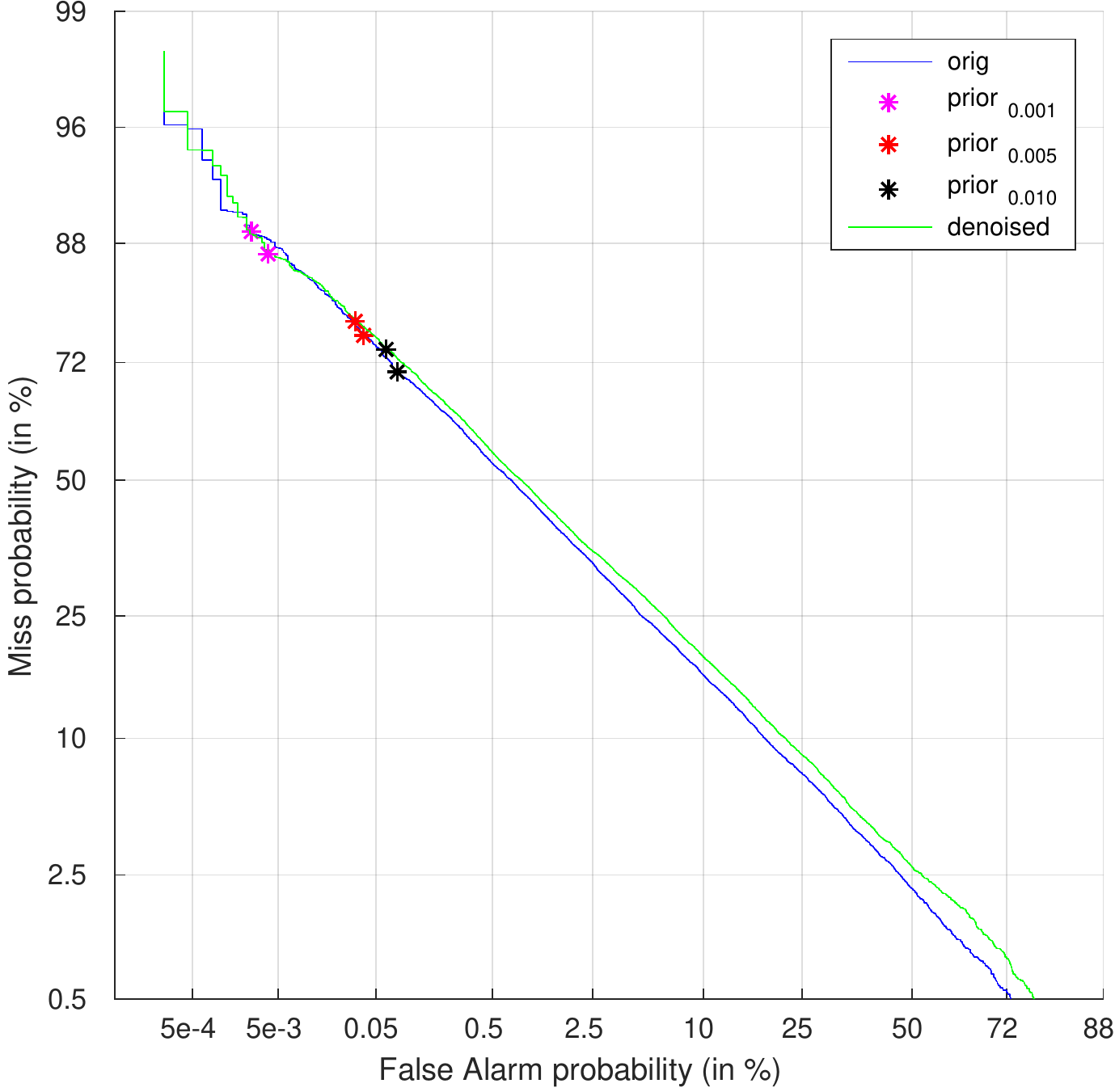}}
    }
    \vspace{1em}
    \centerline{
    \scalebox{1.0}{\includegraphics[width=0.5\linewidth]{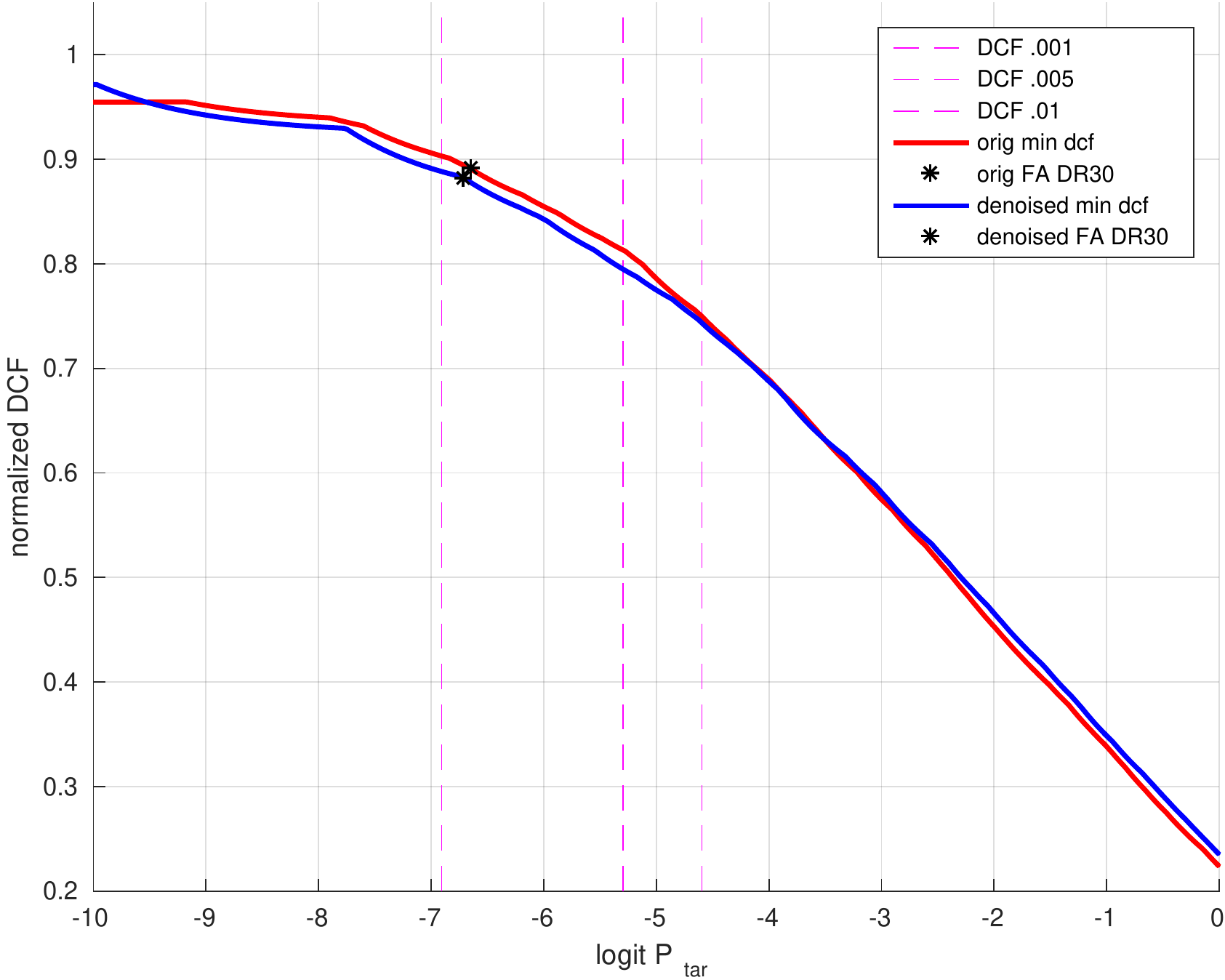}}
    \scalebox{1.0}{\includegraphics[width=0.5\linewidth]{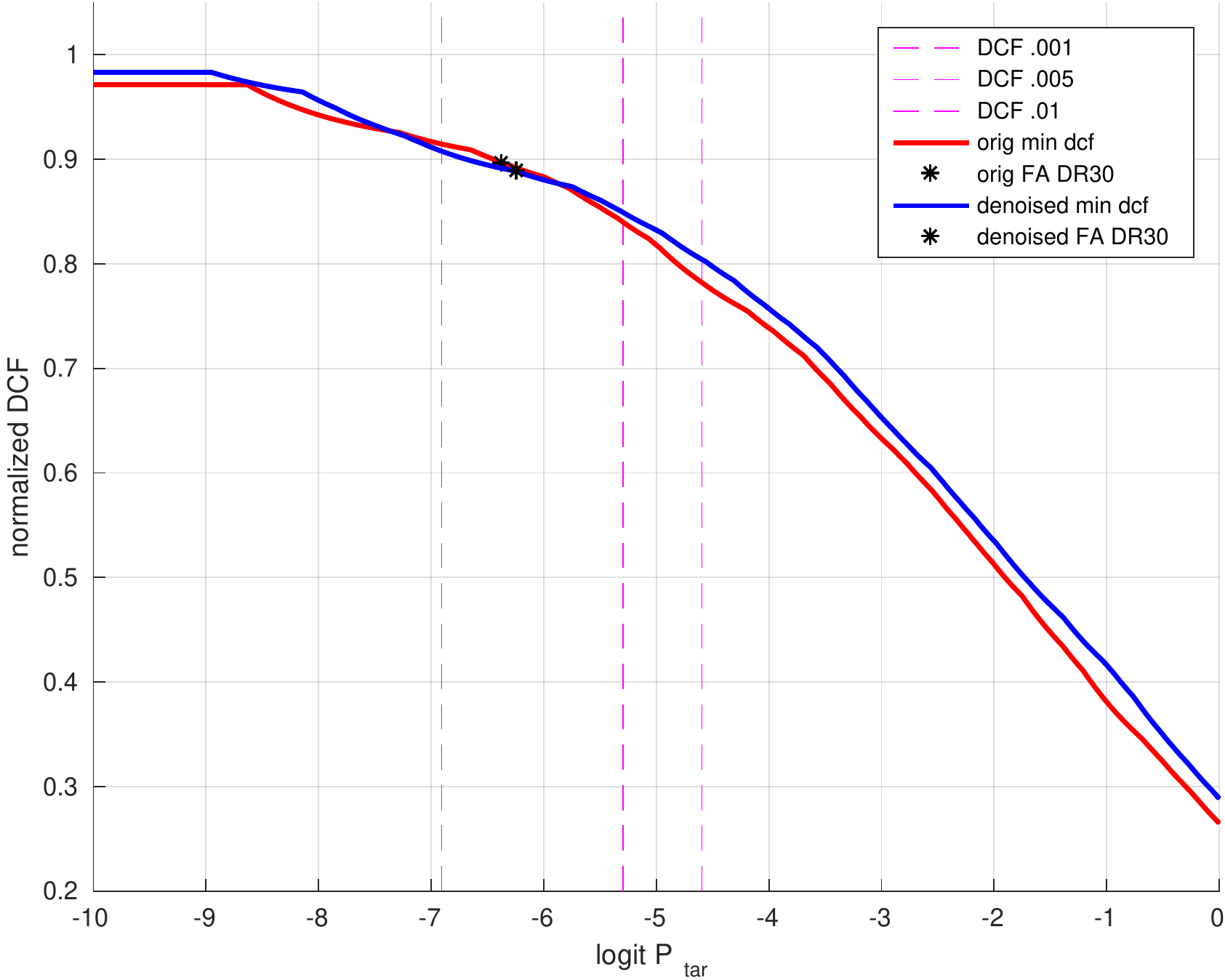}}
}
  \caption{I-vector based systems the left column for MFCC features, the right column for SBN-MFCC features.}
  \label{fig:sre16-yue-f-ivec}
  
\end{subfigure}%
\hspace{0.4em}
\begin{subfigure}{.48\textwidth}
  \centering
    \centerline{
    \scalebox{1.0}{\includegraphics[width=0.5\linewidth]{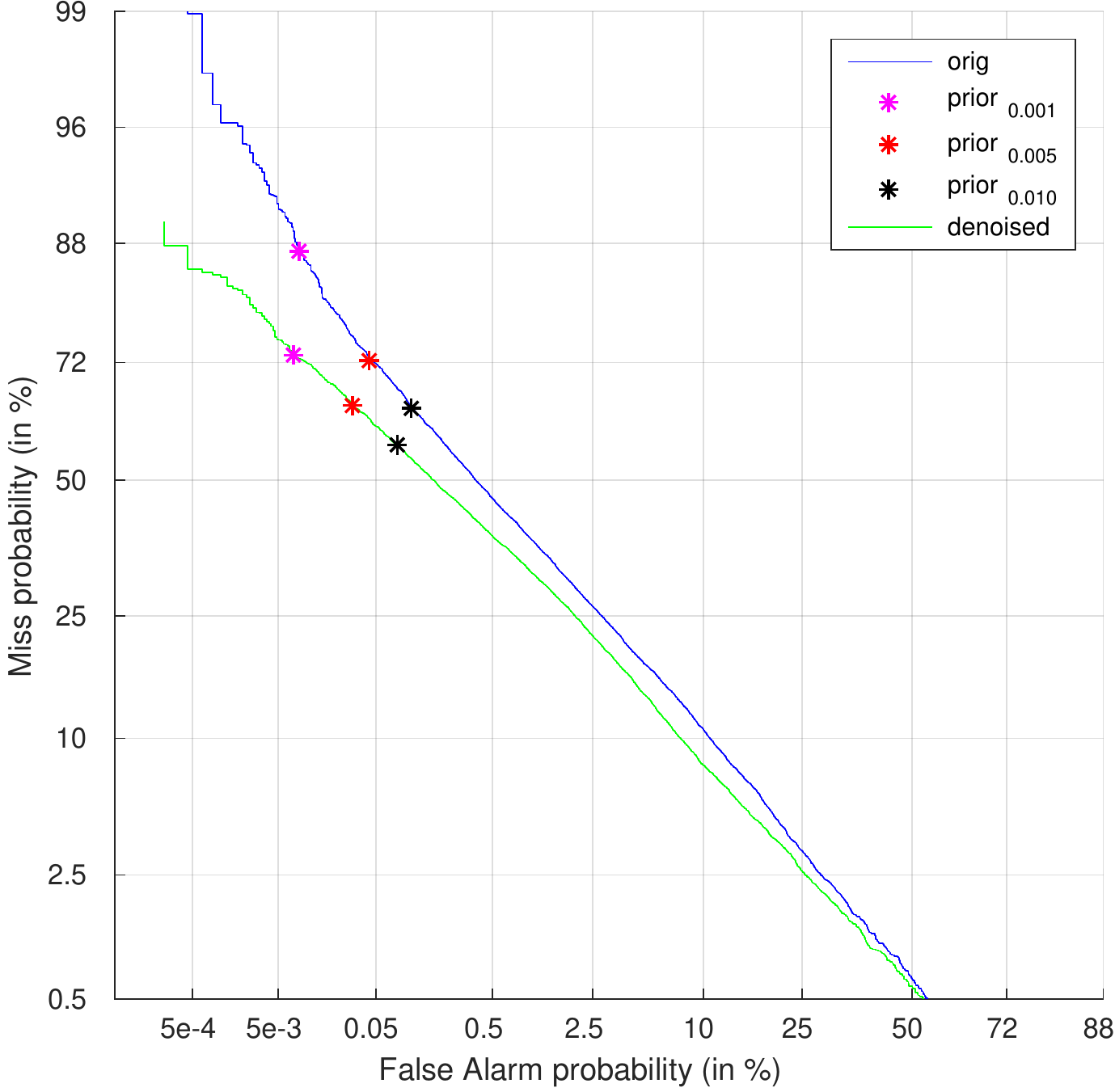}}
    \scalebox{1.0}{\includegraphics[width=0.5\linewidth]{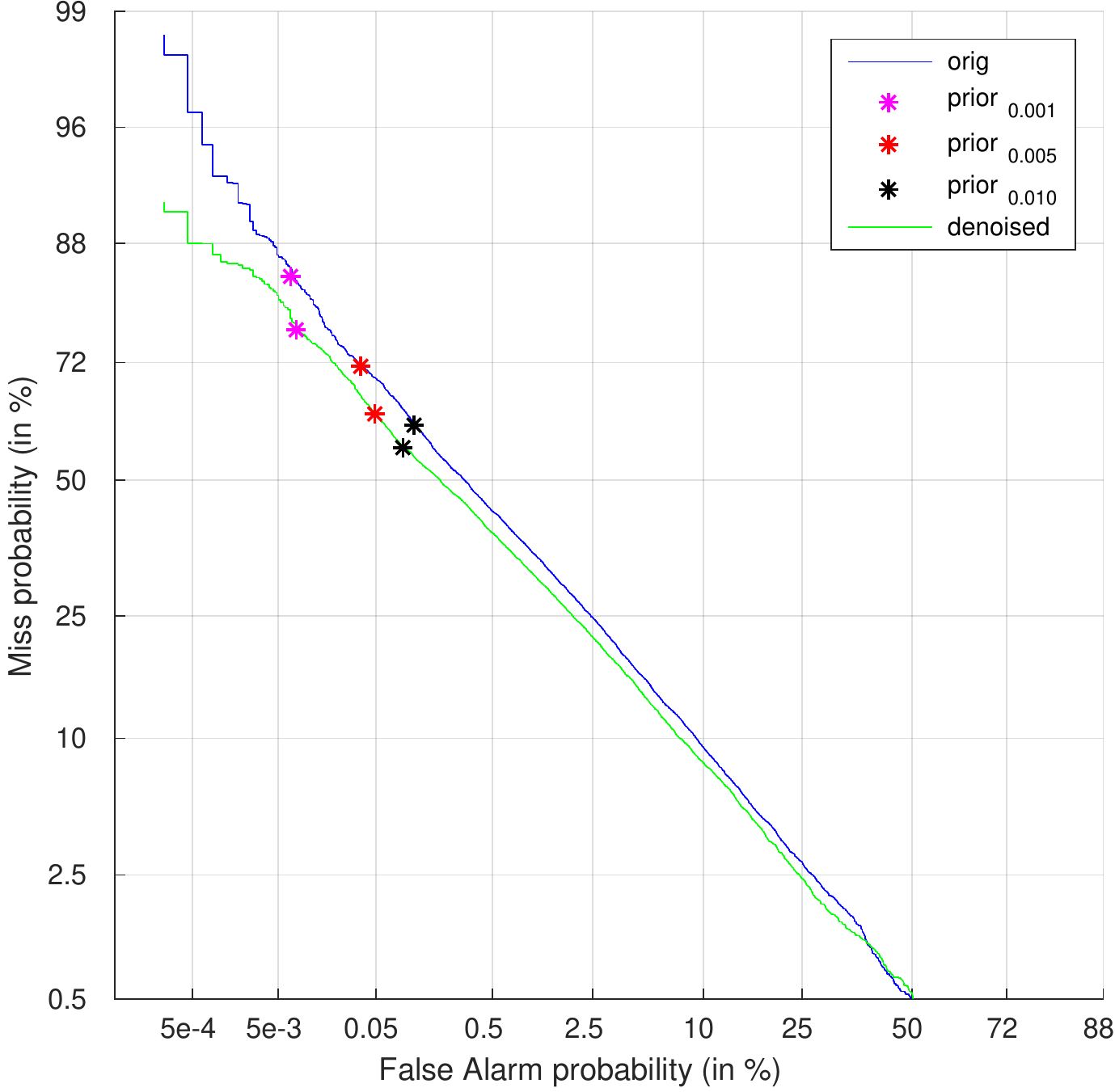}}
    }
    \vspace{1em}
    \centerline{
    \scalebox{1.0}{\includegraphics[width=0.5\linewidth]{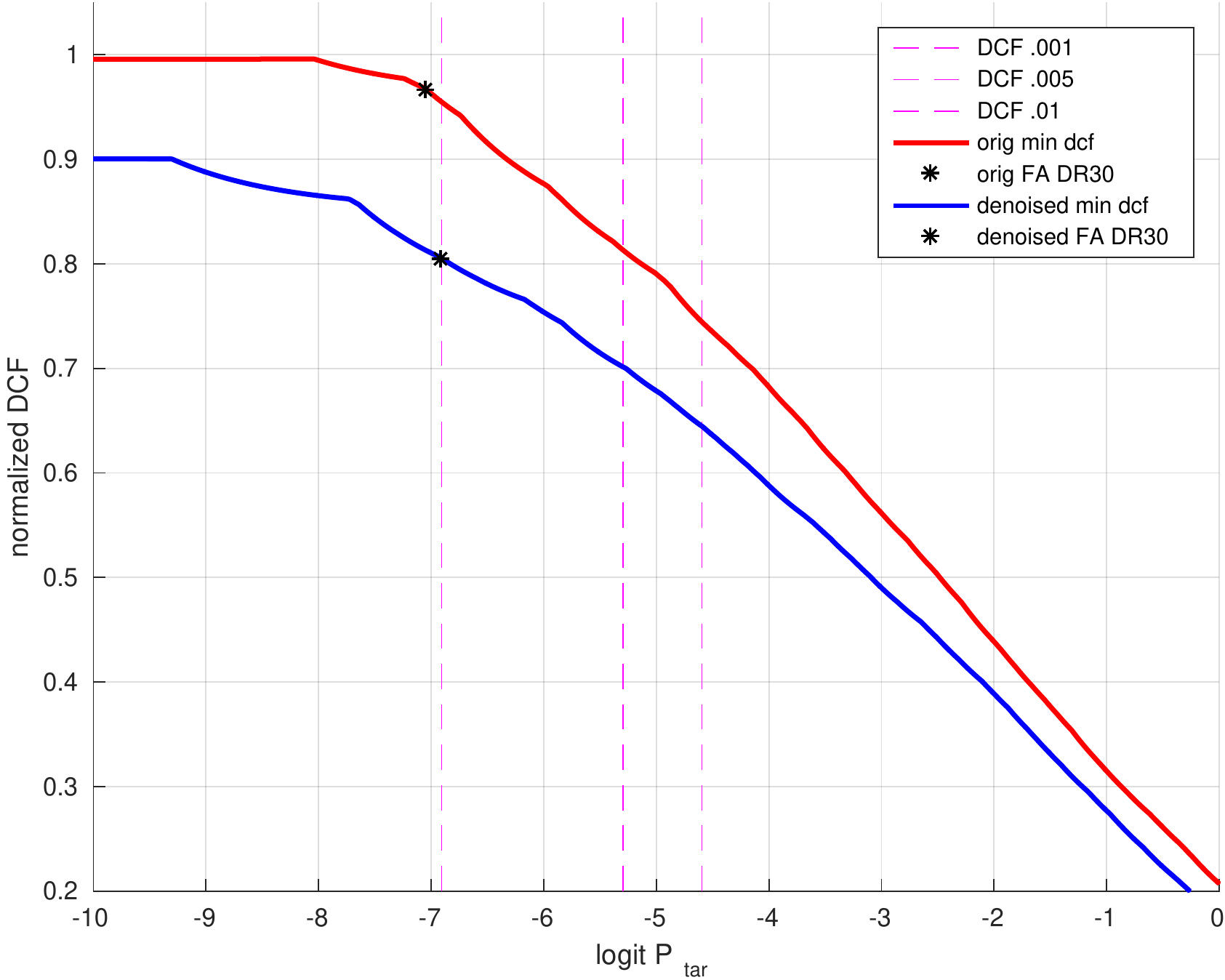}}
    \scalebox{1.0}{\includegraphics[width=0.5\linewidth]{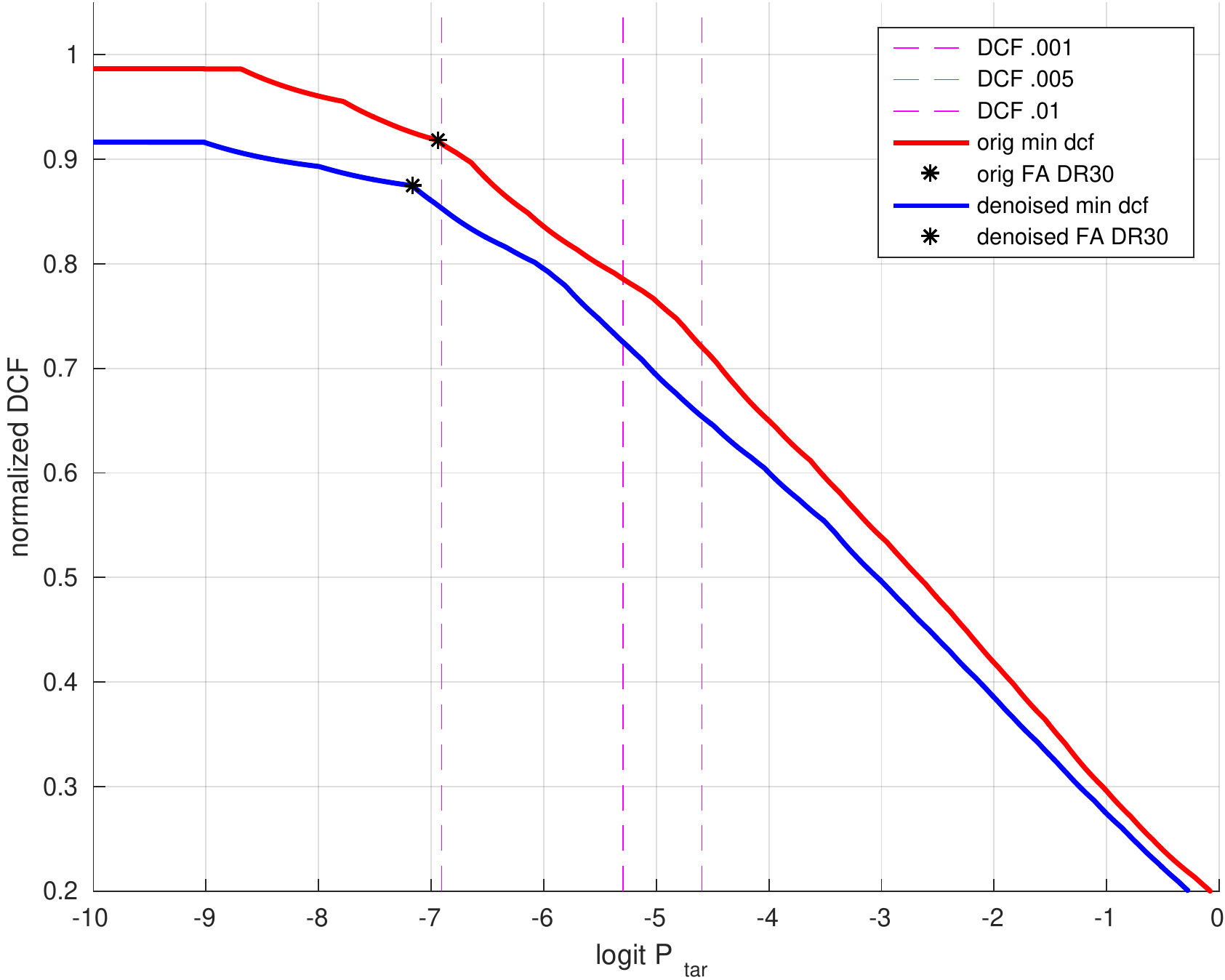}}
    }
  \caption{X-vector based systems the left column for MFCC features, the right column for SBN-MFCC features.}
  \label{fig:sre16-yue-f-xvec}
\end{subfigure}
\caption{ Detection Error Trade-off (DET) plots (top row) and minDFC as a function of effective prior (bottom row) of all tested scenarios for  {\it sre16-yue-f} condition. Intersection of minDCF curves with vertical dashed violet lines correspond from the let to the minDCF from NIST SRE 2010 and to the two operating points of DCF from NIST SRE2016. Similarly the violet star in the DET plots corresponds to the minDCF from NIST SRE2010 and red and black stars correspond to the two operating points of the NIST SRE 2016.}
\label{fig:sre16-yue-f}
\end{figure}
\begin{figure}[p!]
\centering
\begin{subfigure}{.48\textwidth}
  \centering
    \centerline{
    \scalebox{1.0}{\includegraphics[width=0.5\linewidth]{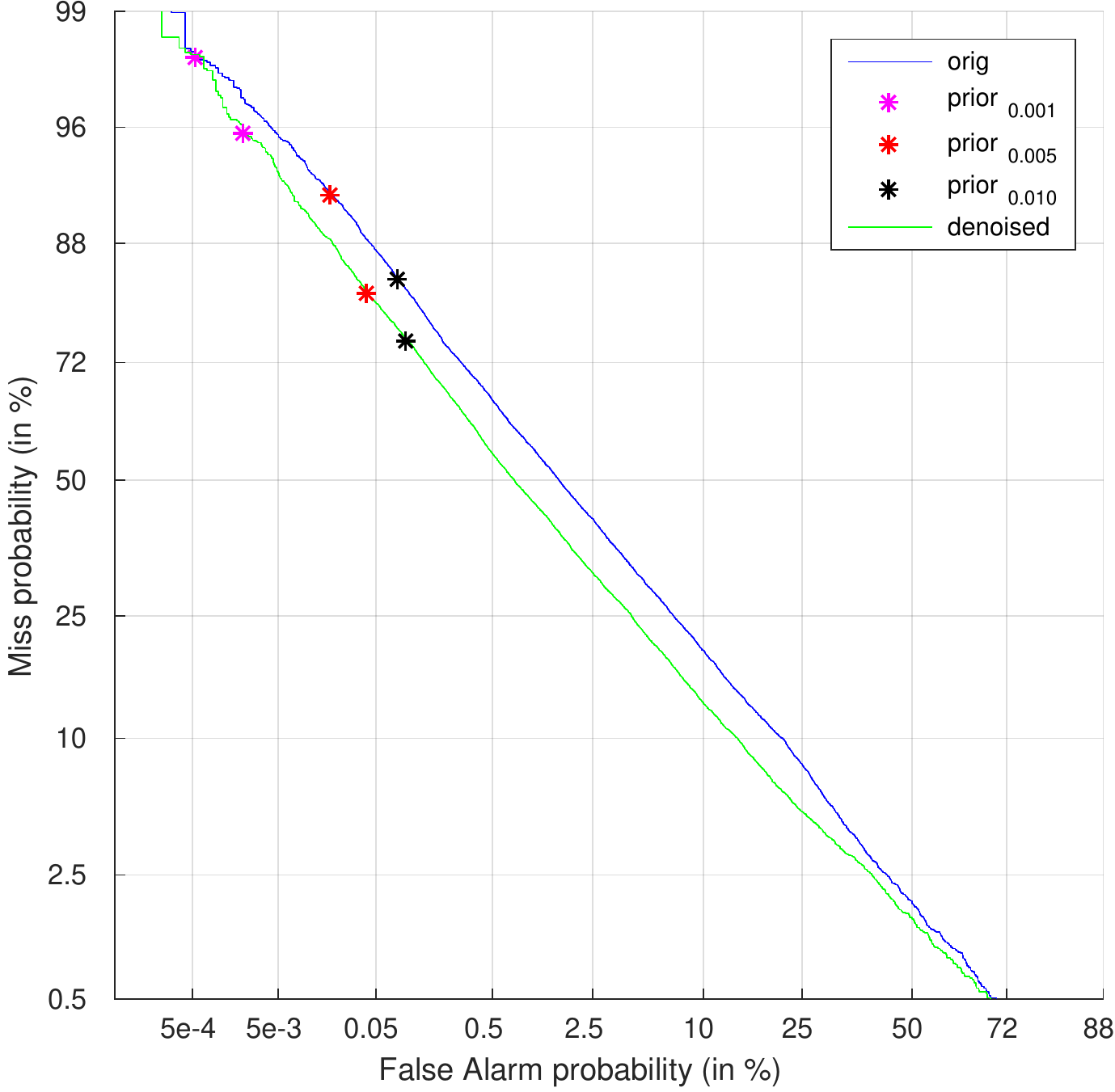}}
    \scalebox{1.0}{\includegraphics[width=0.5\linewidth]{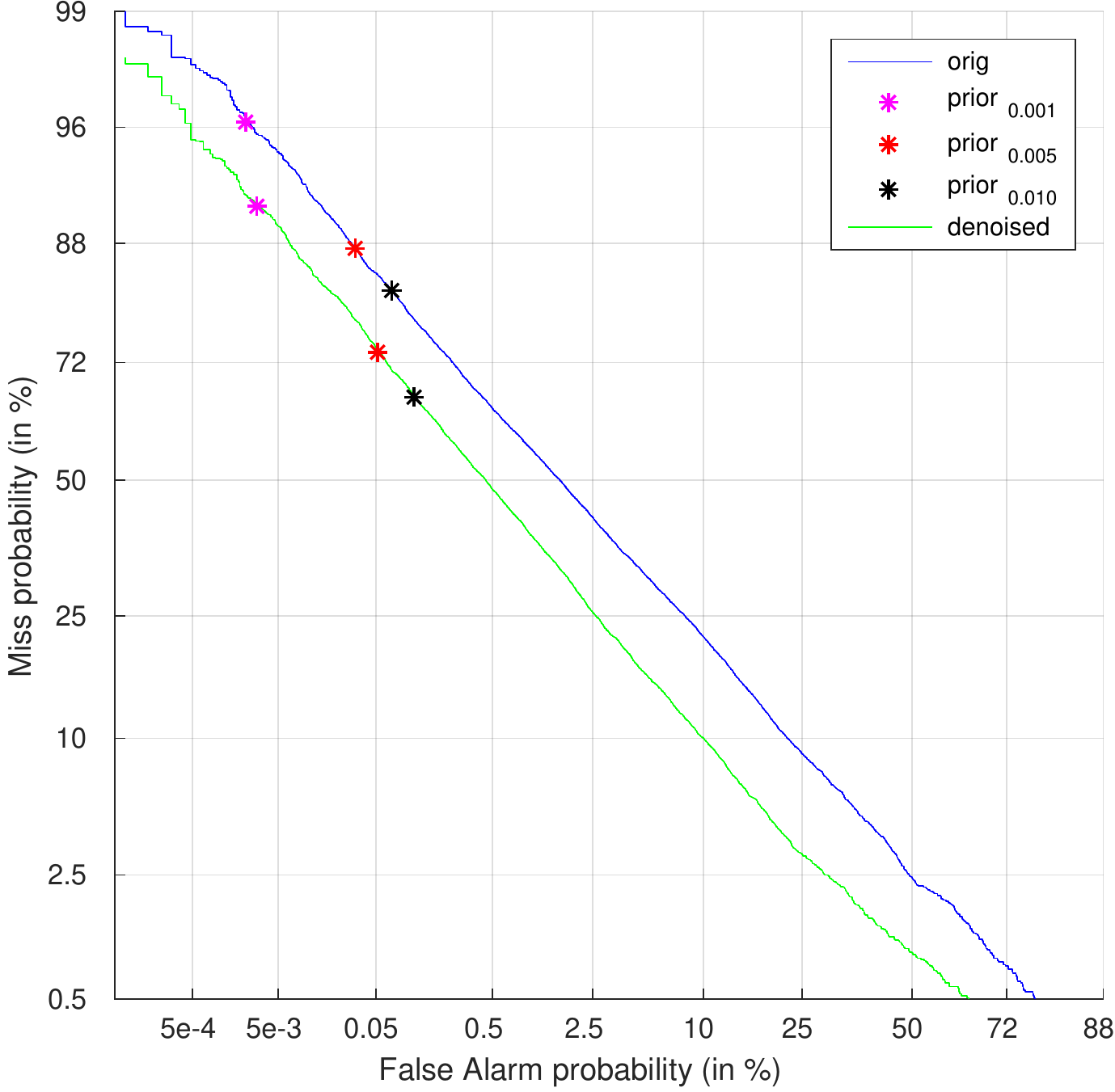}}
    }
    \vspace{1em}
    \centerline{
    \scalebox{1.0}{\includegraphics[width=0.5\linewidth]{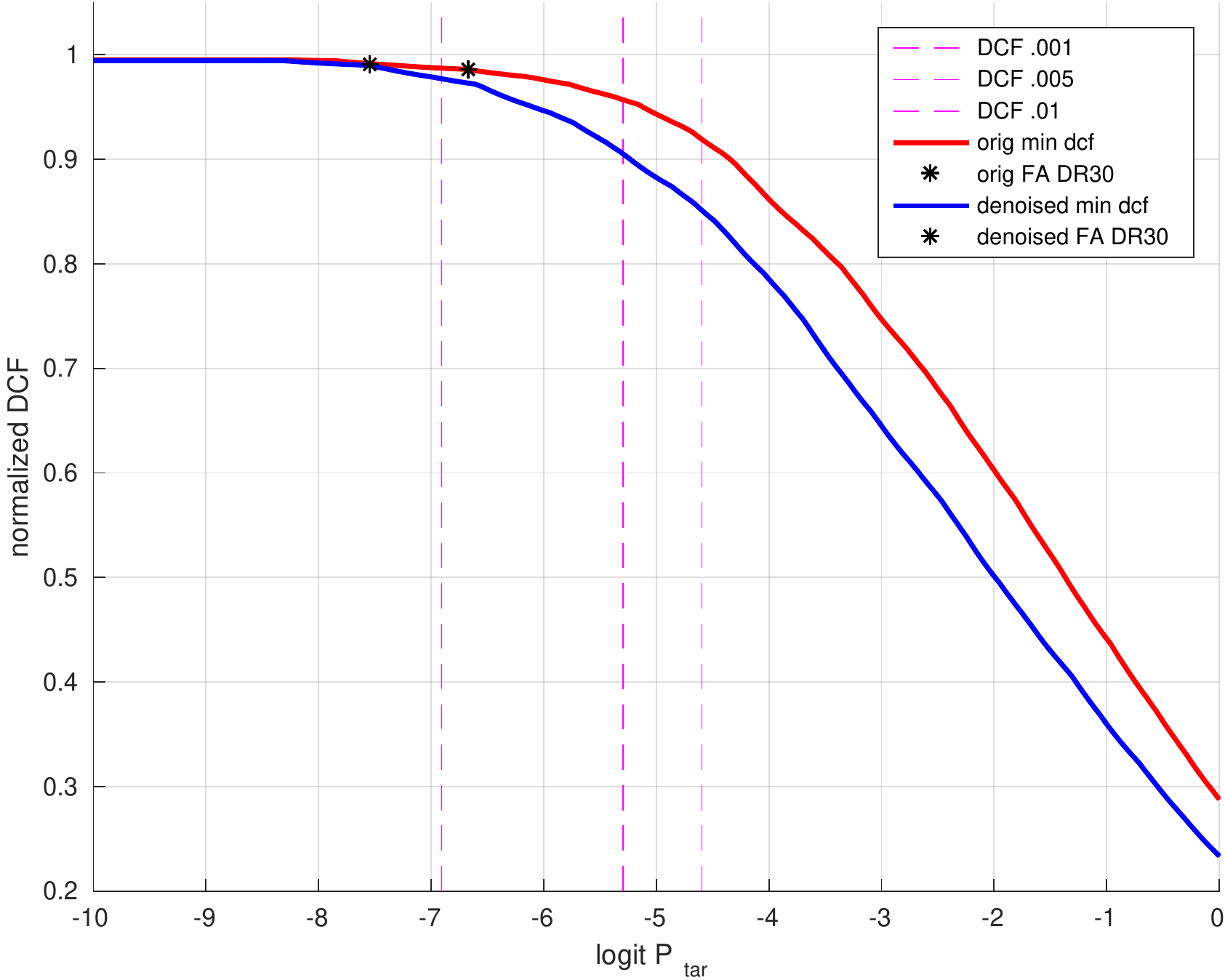}}
    \scalebox{1.0}{\includegraphics[width=0.5\linewidth]{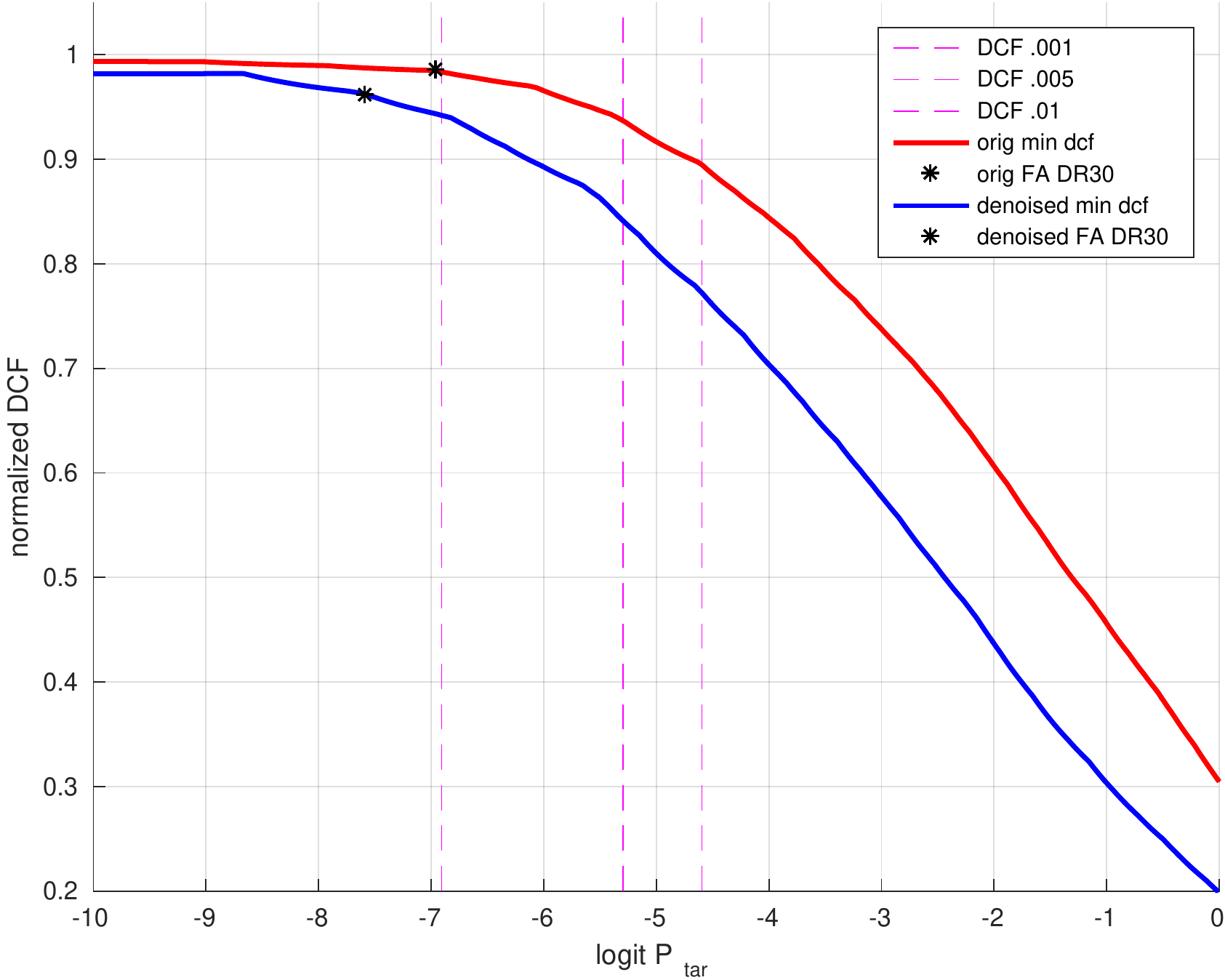}}
}
  \caption{I-vector based systems the left column for MFCC features, the right column for SBN-MFCC features.}
  \label{fig:BUT-RET-merge-ivec}
\end{subfigure}%
\hspace{0.4em}
\begin{subfigure}{.48\textwidth}
  \centering
    \centerline{
    \scalebox{1.0}{\includegraphics[width=0.5\linewidth]{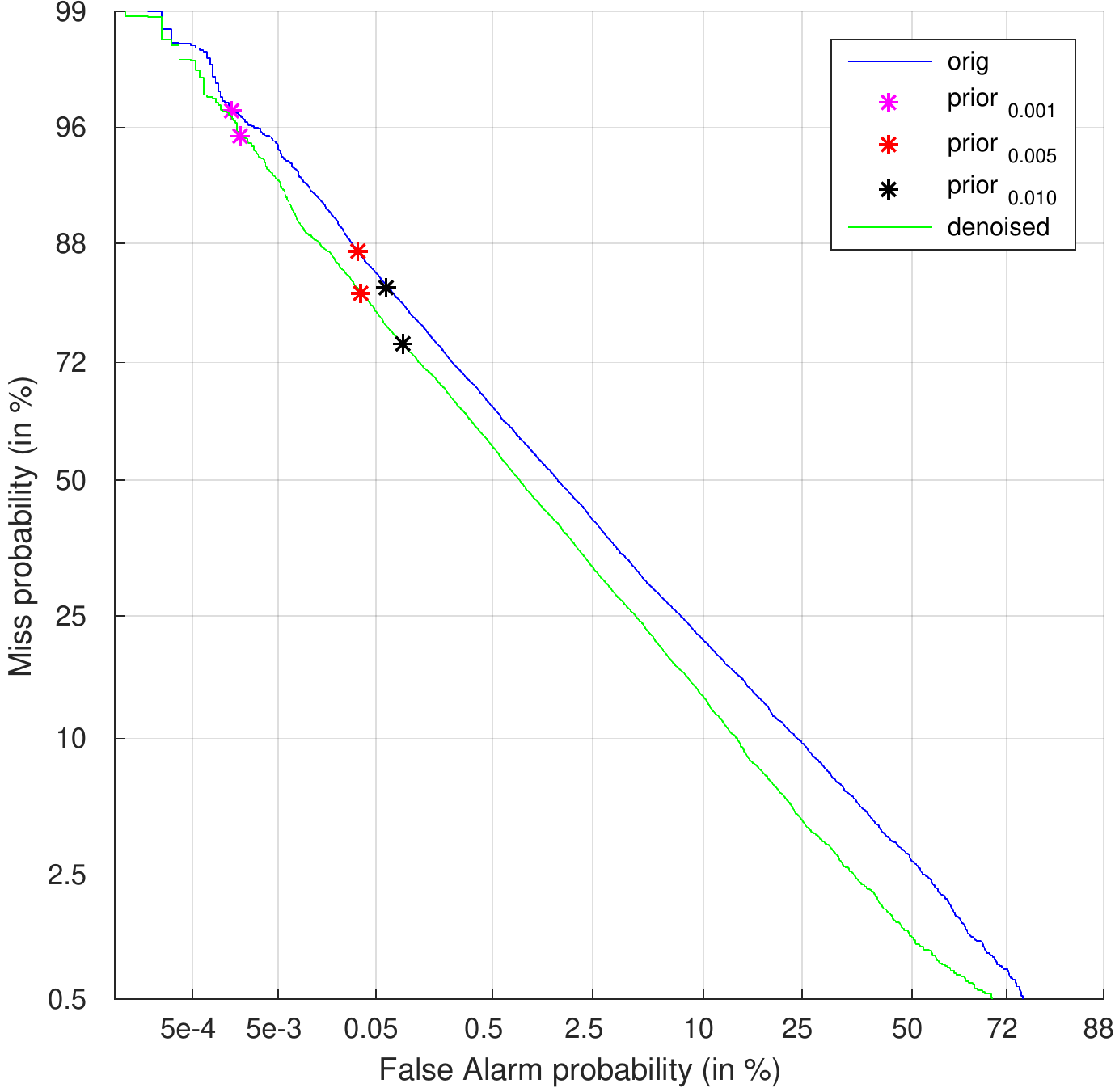}}
    \scalebox{1.0}{\includegraphics[width=0.5\linewidth]{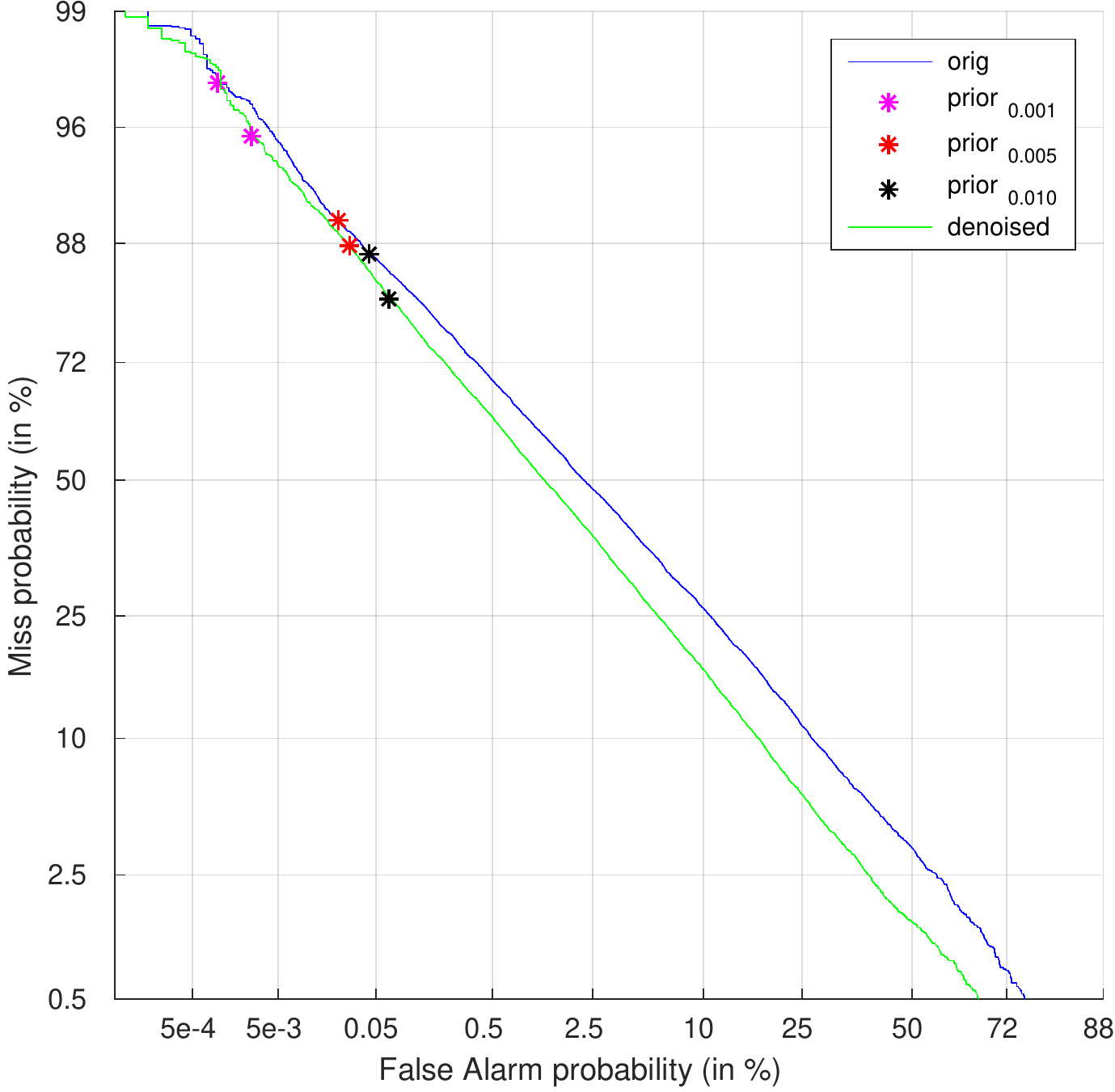}}
    }
    
    \vspace{1em}
    \centerline{
    \scalebox{1.0}{\includegraphics[width=0.5\linewidth]{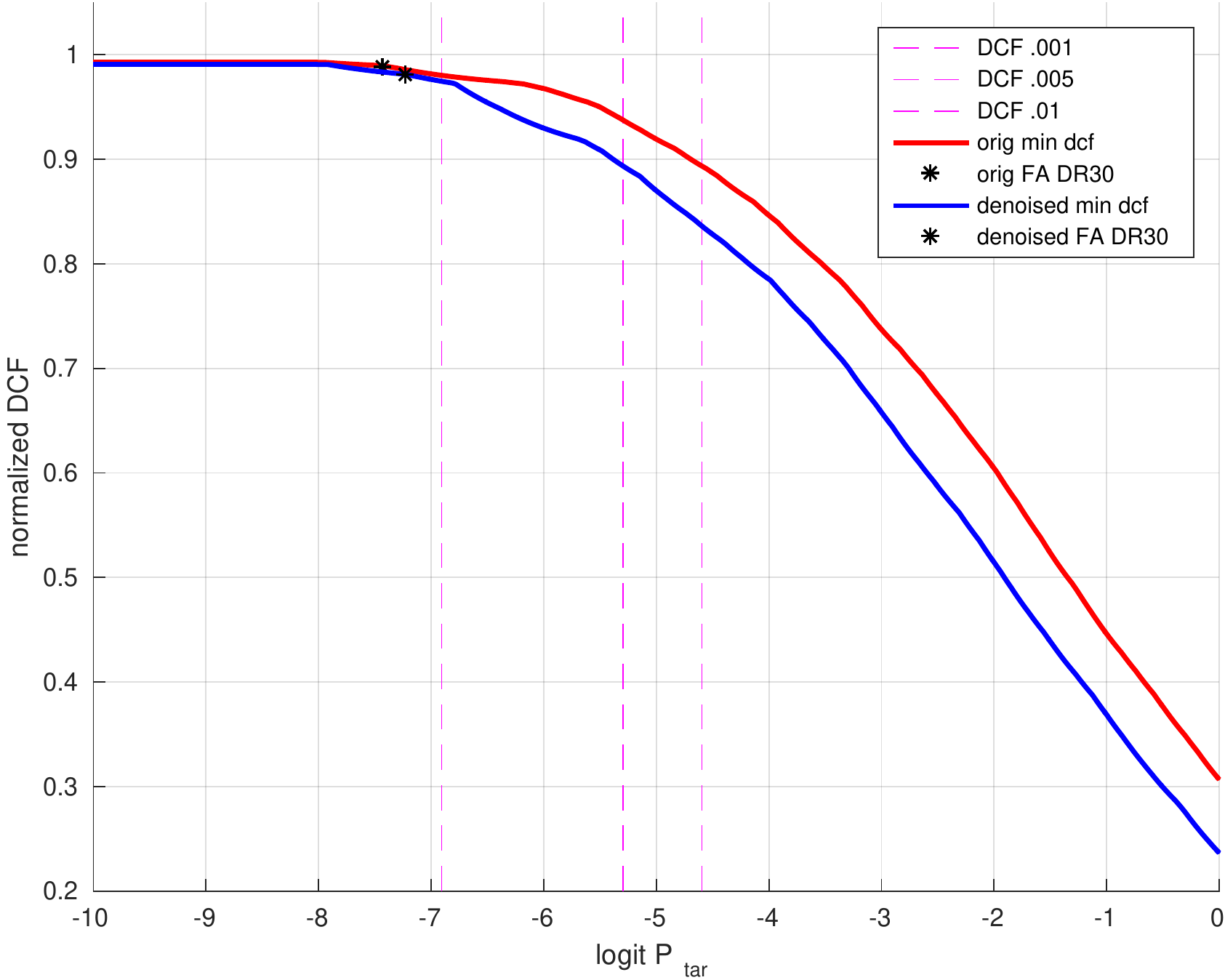}}
    \scalebox{1.0}{\includegraphics[width=0.5\linewidth]{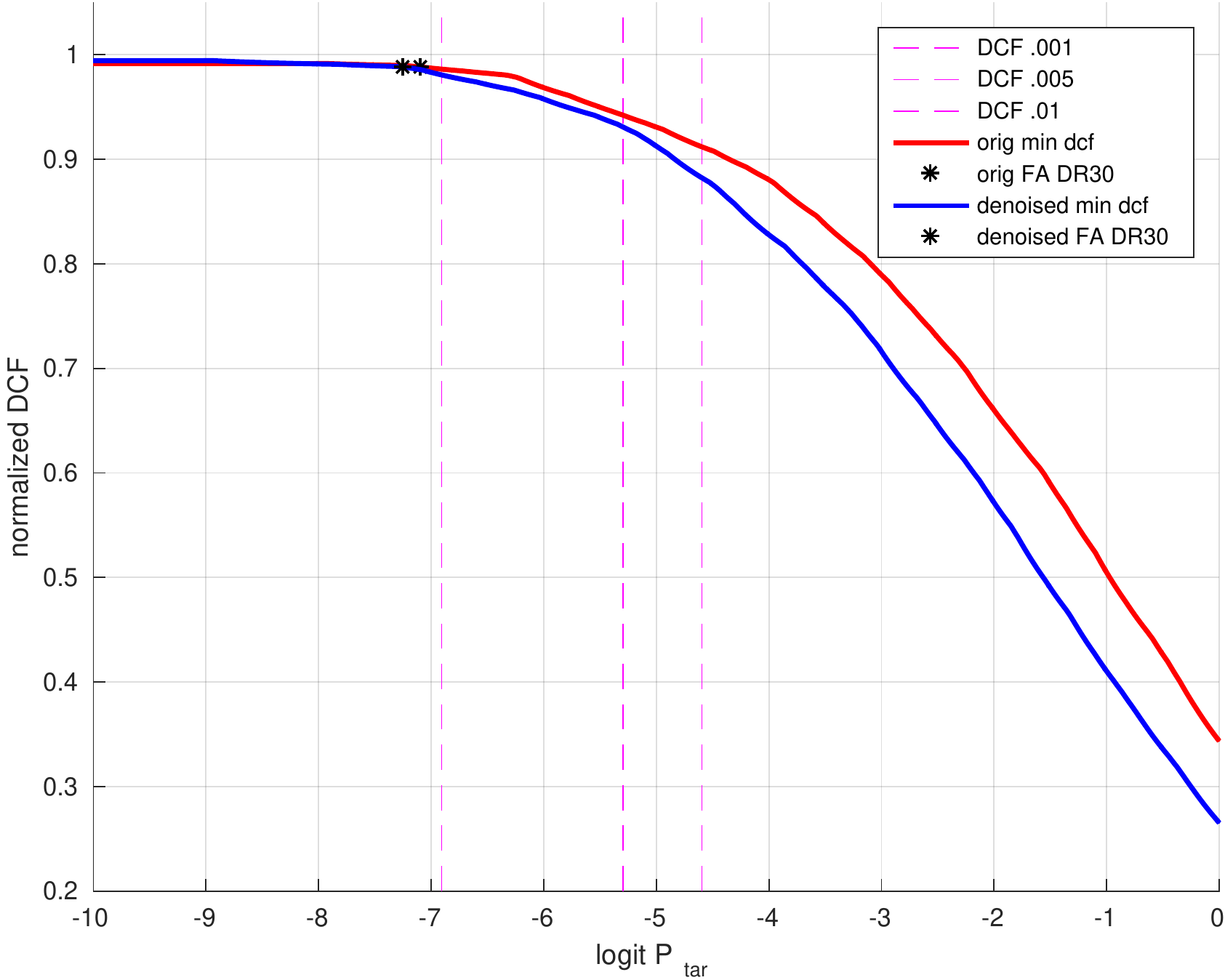}}
    }
  \caption{X-vector based systems the left column for MFCC features, the right column for SBN-MFCC features.}
  \label{fig:BUT-RET-merge-xvec}
\end{subfigure}
\caption{Detection Error Trade-off (DET) plots (top row) and minDFC as a function of effective prior (bottom row) of all tested scenarios for {\it BUT-RET-merge} condition. Intersection of minDCF curves with vertical dashed violet lines correspond from the let to the minDCF from NIST SRE 2010 and to the two operating points of DCF from NIST SRE2016. Similarly the violet star in the DET plots corresponds to the minDCF from NIST SRE2010 and red and black stars correspond to the two operating points of the NIST SRE 2016. }
\label{fig:BUT-RET-merge}  
\end{figure}

Although EER is a common metric summarizing performance, it does not cover all operating points.
In this section, we present the performance of various systems via DET and DCF curves as to see a complex behavior of the systems.

In order to summarize our observation without overwhelming the reader with too many plots, we have chosen two representative conditions, that are closest to the real-world 
scenario---{\it sre16-yue-f}\enspace(Figure~\ref{fig:sre16-yue-f}) and {\it BUT-RET-merge}
(see Figure~\ref{fig:BUT-RET-merge}).  More specifically, the {\it sre16-yue-f} condition
was chosen because a) it contains original noisy audio, and b) compared to the 
rest of the conditions, there is a high channel mismatch between the training 
data and the evaluation data. The {\it BUT-RET-merge} condition was chosen 
because it realistically reflects real reverberation.

Looking at the graphs reveals that the benefit from using the 
studied techniques can be substantial.  It is worth noting that according to the tables 
above,  denoising may not be effective w.r.t. EER, however, when looking at the DET 
curves, we see that there are operating points that do benefit from denoising in a fairly 
large extent.

Apart from i-vector system on the {\it sre16-yue-f} condition, the DET or DCF curves corresponding to the denoised system are generally better than those using the original noisy data over the whole range of operating points.

\section{Conclusion}

In this paper, we analyzed several aspects of DNN-autoencoder enhancement for designing robust speaker verification systems. We studied the influence of the enhancement on different speaker verification system paradigms (generative i-vectors vs. discriminative x-vectors) and we analyzed possible improvement with different features. 

Our results indicate that the DNN autoencoder speech signal enhancement can be helpful to improve system robustness against noise and reverberation. Our results confirm, that it is a stable and universal technique for robustness improvement independently on the system. We also compared the PLDA multi-condition training with audio enhancement.  Both approaches are complementary and systems can benefit from simultaneous usage of both.

After observing improvements achieved with enhancement of the x-vector extractor training data, a possible future work is to train the x-vector extractor in a multi-task fashion, combining speaker separation and signal enhancement objective functions and possibly benefit even more from the joint optimization.

\section*{Acknowledgments}
The work was supported by Czech Ministry of Interior project No. VI20152020025 ``DRAPAK'', Google Faculty Research Award program, Czech Science Foundation under project No. GJ17-23870Y, and by Czech Ministry of Education, Youth and Sports from the National Programme of Sustainability (NPU II) project ``IT4Innovations excellence in science - LQ1602''.





\section*{References}
\bibliographystyle{elsarticle-harv} 
\bibliography{main.bib}

\begin{thebibliography}{53}
\expandafter\ifx\csname natexlab\endcsname\relax\def\natexlab#1{#1}\fi
\expandafter\ifx\csname url\endcsname\relax
  \def\url#1{\texttt{#1}}\fi
\expandafter\ifx\csname urlprefix\endcsname\relax\def\urlprefix{URL }\fi

\bibitem[{Aarts(1992)}]{aweighting}
Aarts, R.~M., 1992. A comparison of {S}ome {L}oudness {M}easures for
  {L}oudspeaker {L}istening {T}ests. J. Audio Eng. Soc 40~(3), 142--146,
  \url{http://www.extra.research.philips.com/hera/people/aarts/RMA_papers/aar92a.pdf}.

\bibitem[{Bhattacharya et~al.(2017)Bhattacharya, Alam, and
  Kenny}]{Bhattacharaya_interspeech_2017}
Bhattacharya, G., Alam, J., Kenny, P., 08 2017. {D}eep {S}peaker {E}mbeddings
  for {S}hort-{D}uration {S}peaker {V}erification. In: Interspeech 2017. pp.
  1517--1521.

\bibitem[{Bhattacharya et~al.(2016)Bhattacharya, Alam, Kenny, and
  Gupta}]{Bhattacharya_SLT16}
Bhattacharya, G., Alam, J., Kenny, P., Gupta, V., 2016. Modelling speaker and
  channel variability using deep neural networks for robust speaker
  verification. In: 2016 {IEEE} Spoken Language Technology Workshop, {SLT}
  2016, San Diego, CA, USA, December 13-16.

\bibitem[{Dehak et~al.(2011)Dehak, Kenny, Dehak, Dumouchel, and
  Ouellet}]{DehakN_TASLP:2010}
Dehak, N., Kenny, P., Dehak, R., Dumouchel, P., Ouellet, P., May 2011.
  Front-{E}nd {F}actor {A}nalysis {F}or {S}peaker {V}erification. IEEE
  Transactions on Audio, Speech, and Language Processing 19~(4), 788--798.

\bibitem[{Dufera and Shimamura(2009)}]{Dufera2009}
Dufera, B., Shimamura, T., Jan 2009. Reverberated speech enhancement using
  neural networks. In: Proc. International Symposium on Intelligent Signal
  Processing and Communication Systems, ISPACS 2009. pp. 441--444.

\bibitem[{ETSI(2007)}]{ETSI:07}
ETSI, 2007. {S}peech {P}rocessing, {T}ransmission and {Q}uality {A}spects
  {(STQ)}. Tech. Rep. ETSI ES 202 050, European Telecommunications Standards
  Institute (ETSI).

\bibitem[{Ferrer et~al.(2011)Ferrer, Bratt, Burget, Cernocky, Glembek,
  Graciarena, Lawson, Lei, Matejka, Plchot, and Scheffer}]{ferrer:sre11}
Ferrer, L., Bratt, H., Burget, L., Cernocky, H., Glembek, O., Graciarena, M.,
  Lawson, A., Lei, Y., Matejka, P., Plchot, O., Scheffer, N., Dec. 2011.
  Promoting robustness for speaker modeling in the community: the {PRISM}
  evaluation set. In: Proceedings of {SRE11} analysis workshop. Atlanta.

\bibitem[{Garcia-Romero and Espy-Wilson(2011)}]{RomeroD_ICSLP:2011}
Garcia-Romero, D., Espy-Wilson, C.~Y., 2011. Analysis of i-vector length
  normalization in {Gaussian}-{PLDA} speaker recognition systems. In: Proc.
  Interspeech.

\bibitem[{Ghahabi and Hernando(2014)}]{Ghahabi_icassp_2014}
Ghahabi, O., Hernando, J., May 2014. Deep belief networks for i-vector based
  speaker recognition. In: 2014 IEEE International Conference on Acoustics,
  Speech and Signal Processing (ICASSP). pp. 1700--1704.

\bibitem[{Glembek et~al.(2014)Glembek, Ma, Mat{\v{e}}jka, Zhang, Plchot,
  Burget, and Matsoukas}]{glembek:domainadaptation}
Glembek, O., Ma, J., Mat{\v{e}}jka, P., Zhang, B., Plchot, O., Burget, L.,
  Matsoukas, S., 2014. {D}omain {A}daptation {V}ia {W}ithin-class {C}ovariance
  {C}orrection in {I}-{V}ector {B}ased {S}peaker {R}ecognition {S}ysterms. In:
  Proceedings of ICASSP 2014. IEEE Signal Processing Society, pp. 4060--4064.
\newline\urlprefix\url{http://www.fit.vutbr.cz/research/view_pub.php?id=10555}

\bibitem[{Heigold et~al.(2016)Heigold, Moreno, Bengio, and
  Shazeer}]{heighold_icassp_2016}
Heigold, G., Moreno, I., Bengio, S., Shazeer, N., March 2016. End-to-end
  text-dependent speaker verification. In: 2016 IEEE International Conference
  on Acoustics, Speech and Signal Processing (ICASSP). pp. 5115--5119.

\bibitem[{{ITU}(1994)}]{telfilter}
{ITU}, October 1994. {ITU-T O.41}.
  \url{https://www.itu.int/rec/dologin_pub.asp?lang=e&id=T-REC-O.41-199410-I!!PDF-E&type=items}.

\bibitem[{Karafi{\'{a}}t et~al.(2014)Karafi{\'{a}}t, Gr{\'{e}}zl, Vesel{\'{y}},
  Hannemann, Sz{\H{o}}ke, and {\v{C}}ernock{\'{y}}}]{Karafiat:IS2014}
Karafi{\'{a}}t, M., Gr{\'{e}}zl, F., Vesel{\'{y}}, K., Hannemann, M.,
  Sz{\H{o}}ke, I., {\v{C}}ernock{\'{y}}, J., 2014. {BUT} 2014 {Babel} system:
  {Analysis} of adaptation in {NN} based systems. In: Interspeech 2014. pp.
  3002--3006.

\bibitem[{Kenny(2010)}]{PLDA:kenny}
Kenny, P., June 2010. Bayesian speaker verification with {H}eavy--{T}ailed
  {P}riors. keynote presentation, Proc. of Odyssey 2010.

\bibitem[{Ko et~al.(2017)Ko, Peddinti, Povey, Seltzer, and
  Khudanpur}]{RIR:Ko2017}
Ko, T., Peddinti, V., Povey, D., Seltzer, M.~L., Khudanpur, S., March 2017. A
  study on data augmentation of reverberant speech for robust speech
  recognition. In: 2017 IEEE International Conference on Acoustics, Speech and
  Signal Processing (ICASSP). pp. 5220--5224.

\bibitem[{Kumatani et~al.(2012)Kumatani, Arakawa, Yamamoto, McDonough, Raj,
  Singh, and Tashev}]{kumatani:micarray:2012}
Kumatani, K., Arakawa, T., Yamamoto, K., McDonough, J., Raj, B., Singh, R.,
  Tashev, I., December 2012. {M}icrophone {A}rray {P}rocessing for {D}istant
  {S}peech {R}ecognition: {T}owards {R}eal-{W}orld {D}eployment. In: APSIPA
  Annual Summit and Conference. Hollywood, CA, USA.

\bibitem[{Laskowski and Edlund(2010)}]{Laskowski:LREC:2010}
Laskowski, K., Edlund, J., may 2010. A {S}nack implementation and {T}cl/{T}k
  {I}nterface to the {F}undamental {F}requency {V}ariation {S}pectrum
  {A}lgorithm. In: Proceedings of the Seventh International Conference on
  Language Resources and Evaluation (LREC'10). Valletta, Malta.

\bibitem[{Lei et~al.(2012)Lei, Burget, Ferrer, Graciarena, and
  Scheffer}]{lei:multistyle}
Lei, Y., Burget, L., Ferrer, L., Graciarena, M., Scheffer, N., 2012. {T}owards
  {N}oise-{R}obust {S}peaker {R}ecognition {U}sing {P}robabilistic {L}inear
  {D}iscriminant {A}nalysis. In: Proceedings of ICASSP. Kyoto, JP.

\bibitem[{Lei et~al.(2014)Lei, Scheffer, Ferrer, and McLaren}]{Lei_icassp_2014}
Lei, Y., Scheffer, N., Ferrer, L., McLaren, M., May 2014. A novel scheme for
  speaker recognition using a phonetically-aware deep neural network. In: 2014
  IEEE International Conference on Acoustics, Speech and Signal Processing
  (ICASSP). pp. 1695--1699.

\bibitem[{Lozano-Diez et~al.(2016)Lozano-Diez, Silnova, Mat{\v{e}}jka, Glembek,
  Plchot, Pe{\v{s}}{\'{a}}n, Burget, and
  Gonzalez-Rodriguez}]{lozano_odyssey_2016}
Lozano-Diez, A., Silnova, A., Mat{\v{e}}jka, P., Glembek, O., Plchot, O.,
  Pe{\v{s}}{\'{a}}n, J., Burget, L., Gonzalez-Rodriguez, J., 2016. {A}nalysis
  and {O}ptimization of {B}ottleneck {F}eatures for {S}peaker {R}ecognition.
  In: Proceedings of Odyssey 2016. Vol. 2016. International Speech
  Communication Association, pp. 352--357.
\newline\urlprefix\url{http://www.fit.vutbr.cz/research/view_pub.php.cz.iso-8859-2?id=11219}

\bibitem[{Mart{\'{i}}nez et~al.(2014)Mart{\'{i}}nez, Burget, Stafylakis, Lei,
  Kenny, and LLeida}]{david:icassp:vts}
Mart{\'{i}}nez, D.~G., Burget, L., Stafylakis, T., Lei, Y., Kenny, P., LLeida,
  E., 2014. {U}nscented {T}ransform {F}or {I}vector-based {N}oisy {S}peaker
  {R}ecognition. In: Proceedings of ICASSP 2014. Florencie, Italy.

\bibitem[{Mat{\v{e}}jka et~al.(2016)Mat{\v{e}}jka, Glembek, Novotn{\'{y}},
  Plchot, Gr{\'{e}}zl, Burget, and {\v{C}}ernock{\'{y}}}]{Matejka:ICASSP:2016}
Mat{\v{e}}jka, P., Glembek, O., Novotn{\'{y}}, O., Plchot, O., Gr{\'{e}}zl, F.,
  Burget, L., {\v{C}}ernock{\'{y}}, J., 2016. {A}nalysis {O}f {DNN}
  {A}pproaches {T}o {S}peaker {I}dentification. In: Proceedings of the 41th
  IEEE International Conference on Acoustics, Speech and Signal Processing
  (ICASSP 2016), 2016. IEEE Signal Processing Society, pp. 5100--5104.
\newline\urlprefix\url{http://www.fit.vutbr.cz/research/view_pub.php?id=11140}

\bibitem[{Mat{\v{e}}jka et~al.(2017)Mat{\v{e}}jka, Novotn{\'{y}}, Plchot,
  Burget, Diez, and {\v{C}}ernock{\'{y}}}]{matejka:snorm}
Mat{\v{e}}jka, P., Novotn{\'{y}}, O., Plchot, O., Burget, L., Diez, M.~S.,
  {\v{C}}ernock{\'{y}}, J., 2017. {A}nalysis of {S}core {N}ormalization in
  {M}ultilingual {S}peaker {R}ecognition. In: Proceedings of Interspeech 2017.
  Vol. 2017. International Speech Communication Association, pp. 1567--1571.
\newline\urlprefix\url{http://www.fit.vutbr.cz/research/view_pub.php?id=11580}

\bibitem[{Mat\v{e}jka et~al.(2006)Mat\v{e}jka, Burget, Schwarz, and
  \v{C}ernock\'{y}}]{matejka-odyssey-2006}
Mat\v{e}jka, P., Burget, L., Schwarz, P., \v{C}ernock\'{y}, J., 2006. {B}rno
  {U}niversity of {T}echnology {S}ystem for {NIST} 2005 {L}anguage
  {R}ecognition {E}valuation. In: Proceedings of Odyssey 2006. San Juan, Puerto
  Rico.

\bibitem[{Mat\v{e}jka et~al.(2014)}]{Matejka:Odyssey2014}
Mat\v{e}jka, P., et~al., 2014. Neural network bottleneck features for language
  identification. In: IEEE Odyssey: The Speaker and Language Recognition
  Workshop. Joensu, Finland.

\bibitem[{McLaren et~al.(2016)McLaren, Ferrer, Castan, and
  Lawson}]{SITW_evaluation_plan}
McLaren, M., Ferrer, L., Castan, D., Lawson, A., 2016. {T}he {S}peakers in the
  {W}ild ({SITW}) {S}peaker {R}ecognition {D}atabase. In: Interspeech 2016. pp.
  818--822.
\newline\urlprefix\url{http://dx.doi.org/10.21437/Interspeech.2016-1129}

\bibitem[{Mimura et~al.(2014)Mimura, Sakai, and Kawahara}]{Mimura2014}
Mimura, M., Sakai, S., Kawahara, T., 2014. Reverberant speech recognition
  combining deep neural networks and deep autoencoders. In: Proc. Reverb
  Challenge Workshop. Florence, Italy.

\bibitem[{Mo\v{s}ner et~al.(2018)Mo\v{s}ner, Mat\v{e}jka, Novotn\'{y}, and
  \v{C}ernock\'{y}}]{mosner:icassp2018}
Mo\v{s}ner, L., Mat\v{e}jka, P., Novotn\'{y}, O., \v{C}ernock\'{y}, J., 2018.
  {D}ereverberation and beamforming in far-field speaker recognition. In:
  Proceedings of ICASSP.

\bibitem[{{NIST}(2010)}]{NIST_SRE:WWW}
{NIST}, 2010. The {NIST} year 2010 {S}peaker {R}ecognition {E}valuation {P}lan.
  \url{
  https://www.nist.gov/sites/default/files/documents/itl/iad/mig/NIST_SRE10_evalplan-r6.pdf}.

\bibitem[{{NIST}(2016)}]{NIST:SRE2016}
{NIST}, 2016. The {NIST} year 2016 {S}peaker {R}ecognition {E}valuation {P}lan.
  \url{
  https://www.nist.gov/sites/default/files/documents/2016/10/\\07/sre16\_eval\_plan\_v1.3.pdf}.

\bibitem[{Novoselov et~al.(2015)Novoselov, Pekhovsky, Kudashev, Mendelev, and
  Prudnikov}]{Novoselov_interspeech_2015}
Novoselov, S., Pekhovsky, T., Kudashev, O., Mendelev, V.~S., Prudnikov, A.,
  Sept 2015. {N}on-linear {PLDA} for i-vector speaker verification. In: 2017
  IEEE International Conference on Acoustics, Speech and Signal Processing
  (ICASSP). pp. 214--218.

\bibitem[{Novotn{\'{y}} et~al.(2018{\natexlab{a}})Novotn{\'{y}}, Mat{\v{e}}jka,
  Plchot, and Glembek}]{nn:Novotny2018}
Novotn{\'{y}}, O., Mat{\v{e}}jka, P., Plchot, O., Glembek, O.,
  2018{\natexlab{a}}. On the use of dnn autoencoder for robust speaker
  recognition. Tech. rep.
\newline\urlprefix\url{http://www.fit.vutbr.cz/research/view_pub.php.cs?id=11855}

\bibitem[{Novotn{\'{y}} et~al.(2018{\natexlab{b}})Novotn{\'{y}}, Plchot,
  Mat{\v{e}}jka, Mo{\v{s}}ner, and Glembek}]{xvec:Novotny2018}
Novotn{\'{y}}, O., Plchot, O., Mat{\v{e}}jka, P., Mo{\v{s}}ner, L., Glembek,
  O., 2018{\natexlab{b}}. On the use of x-vectors for robust speaker
  recognition. In: Proceedings of Odyssey 2018. Vol. 2018. International Speech
  Communication Association, pp. 168--175.
\newline\urlprefix\url{http://www.fit.vutbr.cz/research/view_pub.php?id=11787}

\bibitem[{Peddinti et~al.(2015)Peddinti, Povey, and
  Khudanpur}]{Peddinti:interspeech:2015}
Peddinti, V., Povey, D., Khudanpur, S., 2015. A time delay neural network
  architecture for efficient modeling of long temporal contexts. In:
  {INTERSPEECH} 2015, 16th Annual Conference of the International Speech
  Communication Association, Dresden, Germany, September 6-10, 2015. pp.
  3214--3218.

\bibitem[{Pelecanos and Sridharan(2006)}]{Pelecanos01}
Pelecanos, J., Sridharan, S., 2006. {F}eature {W}arping for {R}obust {S}peaker
  {V}erification. In: Proceedings of Odyssey 2006: The Speaker and Language
  Recognition Workshop. Crete, Greece.

\bibitem[{Plchot et~al.(2016)Plchot, Burget, Aronowitz, and
  Mat{\v{e}}jka}]{plchot-enhancement-2016}
Plchot, O., Burget, L., Aronowitz, H., Mat{\v{e}}jka, P., 2016. {A}udio
  {E}nhancing {W}ith {DNN} {A}utoencoder {F}or {S}peaker {R}ecognition. In:
  Proceedings of the 41th IEEE International Conference on Acoustics, Speech
  and Signal Processing (ICASSP 2016), 2016. IEEE Signal Processing Society,
  pp. 5090--5094.
\newline\urlprefix\url{http://www.fit.vutbr.cz/research/view_pub.php?id=11139}

\bibitem[{Plchot et~al.(2013)Plchot, Matsoukas, Mat\v{e}jka, Dehak, Ma, Cumani,
  Glembek, He\v{r}mansk\'{y}, Mesgarani, Soufifar, Thomas, Zhang, and
  Zhou}]{plchot:rats13}
Plchot, O., Matsoukas, S., Mat\v{e}jka, P., Dehak, N., Ma, J., Cumani, S.,
  Glembek, O., He\v{r}mansk\'{y}, H., Mesgarani, N., Soufifar, M.~M., Thomas,
  S., Zhang, B., Zhou, X., 2013. {D}eveloping {A} {S}peaker {I}dentification
  {S}ystem {F}or {T}he {DARPA} {RATS} {P}roject. In: Proceedings of ICASSP
  2013. Vancouver, CA.

\bibitem[{Prince(2007)}]{prince:iccv:2007}
Prince, S. J.~D., 2007. {P}robabilistic linear discriminant analysis for
  inferences about identity. In: Proc. International Conference on Computer
  Vision (ICCV). Rio de Janeiro, Brazil.

\bibitem[{Ravanelli et~al.(2016)Ravanelli, Svaizer, and Omologo}]{data-sim}
Ravanelli, M., Svaizer, P., Omologo, M., 2016. {R}ealistic {M}ulti-{M}icrophone
  {D}ata {S}imulation for {D}istant {S}peech {R}ecognition. In: Interspeech
  2016. pp. 2786--2790.
\newline\urlprefix\url{http://dx.doi.org/10.21437/Interspeech.2016-731}

\bibitem[{Rohdin et~al.(2018)Rohdin, Silnova, Diez, Plchot, Mat\v{e}jka, and
  Burget}]{rohdin:icassp:2018}
Rohdin, J., Silnova, A., Diez, M., Plchot, O., Mat\v{e}jka, P., Burget, L.,
  2018. {E}nd-to-end {DNN} based speaker recognition inspired by i-vector and
  {PLDA}. In: Proceedings of ICASSP. IEEE Signal Processing Society.

\bibitem[{Snyder(2017)}]{kaldi:xvec}
Snyder, D., 2017. {NIST} {SRE} 2016 {X}vector {R}ecipe.
  \url{https://david-ryan-snyder.github.io/2017/10/04/model_sre16_v2.html}.

\bibitem[{Snyder et~al.(2017)Snyder, Garcia-Romero, Povey, and
  Khudanpur}]{xvec:Snyder2016}
Snyder, D., Garcia-Romero, D., Povey, D., Khudanpur, S., 2017. {D}eep {N}eural
  {N}etwork {E}mbeddings for {T}ext-{I}ndependent {S}peaker {V}erification.
  Proc. Interspeech 2017, 999--1003.

\bibitem[{Snyder et~al.(2018)Snyder, Garcia-Romero, Sell, Povey, and
  Khudanpur}]{xvec:Snyder2018}
Snyder, D., Garcia-Romero, D., Sell, G., Povey, D., Khudanpur, S., 2018.
  {X}-vectors: {R}obust {DNN} {E}mbeddings for {S}peaker {R}ecognition. In:
  Proceedings of ICASSP.

\bibitem[{Snyder et~al.(2016)Snyder, Ghahremani, Povey, Garcia-Romero, Carmiel,
  and Khudanpur}]{sreEndEnd:Snyder2016}
Snyder, D., Ghahremani, P., Povey, D., Garcia-Romero, D., Carmiel, Y.,
  Khudanpur, S., Dec 2016. Deep neural network-based speaker embeddings for
  end-to-end speaker verification. In: 2016 IEEE Spoken Language Technology
  Workshop (SLT). pp. 165--170.

\bibitem[{Stewart and Sandler(2010)}]{C4DM:WWW2}
Stewart, R., Sandler, M., March 2010. {D}atabase of omnidirectional and
  {B}-format room impulse responses. In: 2010 IEEE International Conference on
  Acoustics, Speech and Signal Processing. pp. 165--168.

\bibitem[{Sturim and Reynolds(2005)}]{Sturim:snorm}
Sturim, D.~E., Reynolds, D.~A., March 2005. {S}peaker adaptive cohort selection
  for {T}norm in text-independent speaker verification. In: Proceedings.
  (ICASSP '05). IEEE International Conference on Acoustics, Speech, and Signal
  Processing, 2005. Vol.~1. pp. I/741--I/744 Vol. 1.

\bibitem[{Swart and Br{\"{u}}mmer(2017)}]{niko:norm}
Swart, A., Br{\"{u}}mmer, N., 2017. {A} {G}enerative {M}odel for {S}core
  {N}ormalization in {S}peaker {R}ecognition. In: Interspeech 2017, 18th Annual
  Conference of the International Speech Communication Association, Stockholm,
  Sweden, August 20-24, 2017. pp. 1477--1481.
\newline\urlprefix\url{http://www.isca-speech.org/archive/Interspeech\_2017/abstracts/0137.html}

\bibitem[{Talkin(1995)}]{talkin:pitch:95}
Talkin, D., 1995. {A} {R}obust {A}lgorithm for {P}itch {T}racking ({RAPT}). In:
  Kleijn, W.~B., Paliwal, K. (Eds.), Speech Coding and Synthesis. Elsevier, New
  York.

\bibitem[{Variani et~al.(2014)Variani, Lei, McDermott, Moreno, and
  Gonzalez-Dominguez}]{Variani_icassp_2014}
Variani, E., Lei, X., McDermott, E., Moreno, I.~L., Gonzalez-Dominguez, J., May
  2014. Deep neural networks for small footprint text-dependent speaker
  verification. In: 2014 IEEE International Conference on Acoustics, Speech and
  Signal Processing (ICASSP). pp. 4052--4056.

\bibitem[{Xu et~al.(2014{\natexlab{a}})Xu, Du, Dai, and Lee}]{Xu2014}
Xu, Y., Du, J., Dai, L.-R., Lee, C.-H., Jan. 2014{\natexlab{a}}. {A}n
  {E}xperimental {S}tudy on {S}peech {E}nhancement {B}ased on {D}eep {N}eural
  {N}etworks. IEEE Signal processing letters 21~(1).

\bibitem[{Xu et~al.(2014{\natexlab{b}})Xu, Du, Dai, and Lee}]{Xu2014a}
Xu, Y., Du, J., Dai, L.-R., Lee, C.-H., 2014{\natexlab{b}}. Global variance
  equalization for improving deep neural network based speech enhancement. In:
  Proc. IEEE China Summit \& International Conference on Signal and Information
  Processing (ChinaSIP). pp. 71 -- 75.

\bibitem[{Yanhui et~al.(2014)Yanhui, Jun, Yong, Lirong, and
  Chin-Hui}]{Yanhui2014}
Yanhui, T., Jun, D., Yong, X., Lirong, D., Chin-Hui, L., 2014. {D}eep {N}eural
  {N}etwork {B}ased {S}peech {S}eparation for {R}obust {S}peech {R}ecognition.
  In: Proceedings of ICSP2014. pp. 532--536.

\bibitem[{Zhang et~al.(2016)Zhang, Chen, Zhao, Li, and Gong}]{zhang_slt_2016}
Zhang, S.~X., Chen, Z., Zhao, Y., Li, J., Gong, Y., Dec 2016. {E}nd-to-{E}nd
  attention based text-dependent speaker verification. In: 2016 IEEE Spoken
  Language Technology Workshop (SLT). pp. 171--178.

\end{thebibliography}





\end{document}